\documentclass[twocolumn,superscriptaddress,floatfix,
longbibliography,aps,pra,nopreprintnumbers,10pt]{revtex4-1}

\usepackage[utf8]{inputenc}
\usepackage[english]{babel}

\usepackage{graphicx}
\usepackage{bm}
\usepackage{xcolor}
\usepackage{epstopdf}
\usepackage{amsmath}
\usepackage{amssymb}
\usepackage{epstopdf}
\usepackage[normalem]{ulem} %strikethrough \sout
\usepackage{enumerate}

\usepackage[urlcolor=blue,colorlinks=true,citecolor=blue,linkcolor=blue,pdfstartview={FitH},bookmarks=false]{hyperref}
\newcommand{\e}{\textrm{e}}
\newcommand{\im}{\textrm{i}}

\newcommand{\bsig}{\bm{\sigma}}
\newcommand{\btau}{\bm{\tau}}

\newcommand{\bs}{\boldsymbol}
\newcommand{\bpm}{\begin{pmatrix}}
\newcommand{\epm}{\end{pmatrix}}

\DeclareMathOperator{\Tr}{tr}

\DeclareMathOperator{\sgn}{sgn}

\begin{document}

\title{Instability of Majorana states in Shiba chains due to leakage into a topological substrate}

\author{Nicholas Sedlmayr}
\email[e-mail: ]{sedlmayr@umcs.pl}
\affiliation{Institute of Physics, Maria Curie-Sk\l{}odowska University,
Plac Marii Sk\l{}odowskiej-Curie 1, PL-20031 Lublin, Poland}
\author{Cristina Bena}
\affiliation{Universit\'e Paris Saclay, CNRS, CEA, Institut de Physique Th\'eorique, 91191, Gif-sur-Yvette, France}

\date{\today}

\begin{abstract}
We revisit the problem of Majorana states in chains of scalar impurities deposited on a superconductor with a mixed s-wave and p-wave pairing. We also study the formation of Majorana states for magnetic impurity chains. We find that the magnetic impurity chains exhibit well-localized Majorana states when the substrate is trivial, but these states hybridize and get dissolved in the bulk when the substrate is topological. Most surprisingly, and contrary to previous predictions, the scalar impurity chain does not support fully localized Majorana states except for very small and finely tuned parameter regimes, mostly for a  non-topological substrate close to the topological transition. Our results indicate that a purely p-wave or a dominant p-wave substrate are not good candidates to support either magnetic or scalar impurity topological Shiba chains.
\end{abstract}

\maketitle

\section{Introduction}
\label{intro}

The presence of impurities in superconductors may give rise under certain conditions to impurity Yu-Shiba-Rusinov states \cite{Yu1965,Shiba1968,Rusinov1969,Yazdani1997,Menard2015}.  There have been many proposals in the past years to use chains of such bound states to create topological Majorana zero modes (MZMs)~\cite{Nadj-Perge2013,Pientka2013,Braunecker2013,Klinovaja2013,Vazifeh2013,Pientka2014,Poyhonen2014,Reis2014,Kim2014,Li2014,Heimes2014,Brydon2015,
Weststrom2015,Heimes2015,Peng2015,Hui2015,Rontynen2015,Braunecker2015,Ebisu2015a,Poyhonen2016,Zhang2016,
Li2016a,Rontynen2016,Hoffman2016,Li2016b,Schecter2016,Christensen2016,Kaladzhyan2017a,
Andolina2017,Kobialka2020,Menard2017,Menard2019,Nadj-Perge2014,Pawlak2015,Ruby2015,Feldman2017a,Ruby2017,Kim2018,Pawlak2019}. 
One of the most common configurations believed to support Majorana states is a chain of scalar impurities deposited on top of a p-wave superconductor (SC)~\cite{Neupert2016,Sahlberg2017,Kaladzhyan2016,Kaladzhyan2018,Kreisel2021}. While to date no material has been demonstrates to be a p-wave superconductor, many candidates have been proposed, the most cited being $\rm Sr_2 Ru O_4$, and there are many others are being currently explored.

Recent proposals indicate that the formation of Majorana states is possible even in configurations in which the substrate is a superconductor with mixed s-wave and p-wave components~\cite{Neupert2016,Kaladzhyan2018}. Mixed spin-singlet and spin-triplet pairing was observed in non-centrosymmetric heavy fermion superconductors, like $\rm CePt_3Si$~\cite{Yanase2005,Hayashi2008a}, and can be envisaged either by coupling a spin-singlet superconductor to a ferromagnet or by inducing superconductivity via proximity in a magnetic material in which, for example, Rashba spin orbit coupling plays an important role~\cite{Reeg2015}. A system in a mixed s-wave/p-wave SC state can be in either a topologically trivial or non-trivial phase, depending on the ratio of the s-wave and p-wave components \cite{Kaladzhyan2018}. It has been proposed that a chain of scalar (non-magnetic) impurities deposited on such a substrate may become topological and exhibit Majorana states for both the regime in which the substrate is topological, as well as for a small region close to the phase transition for which the substrate is topologically trivial \cite{Neupert2016}.

\begin{figure}[t!]
  \includegraphics[width=0.9\columnwidth]{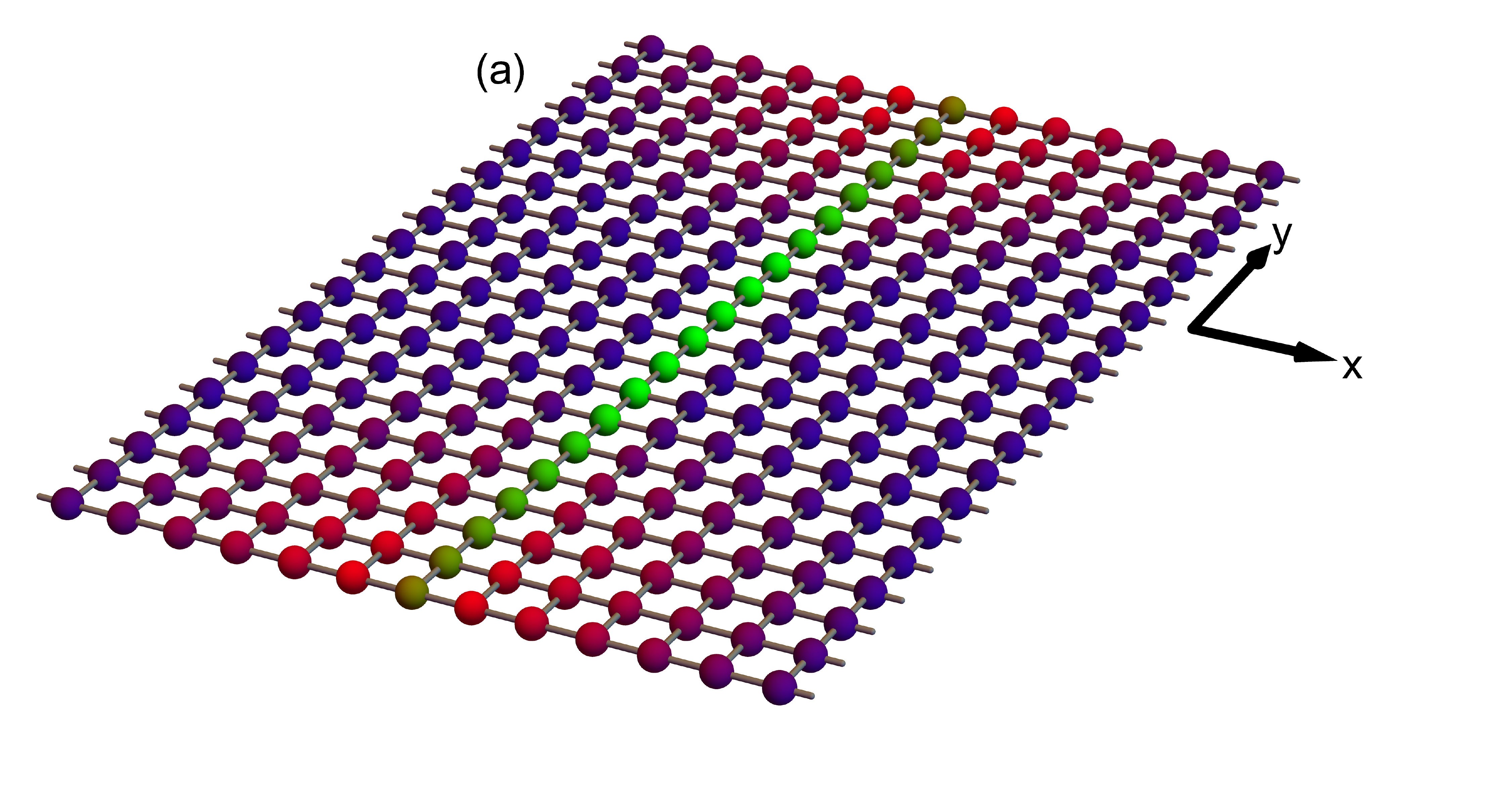}\\
  \includegraphics[width=0.9\columnwidth]{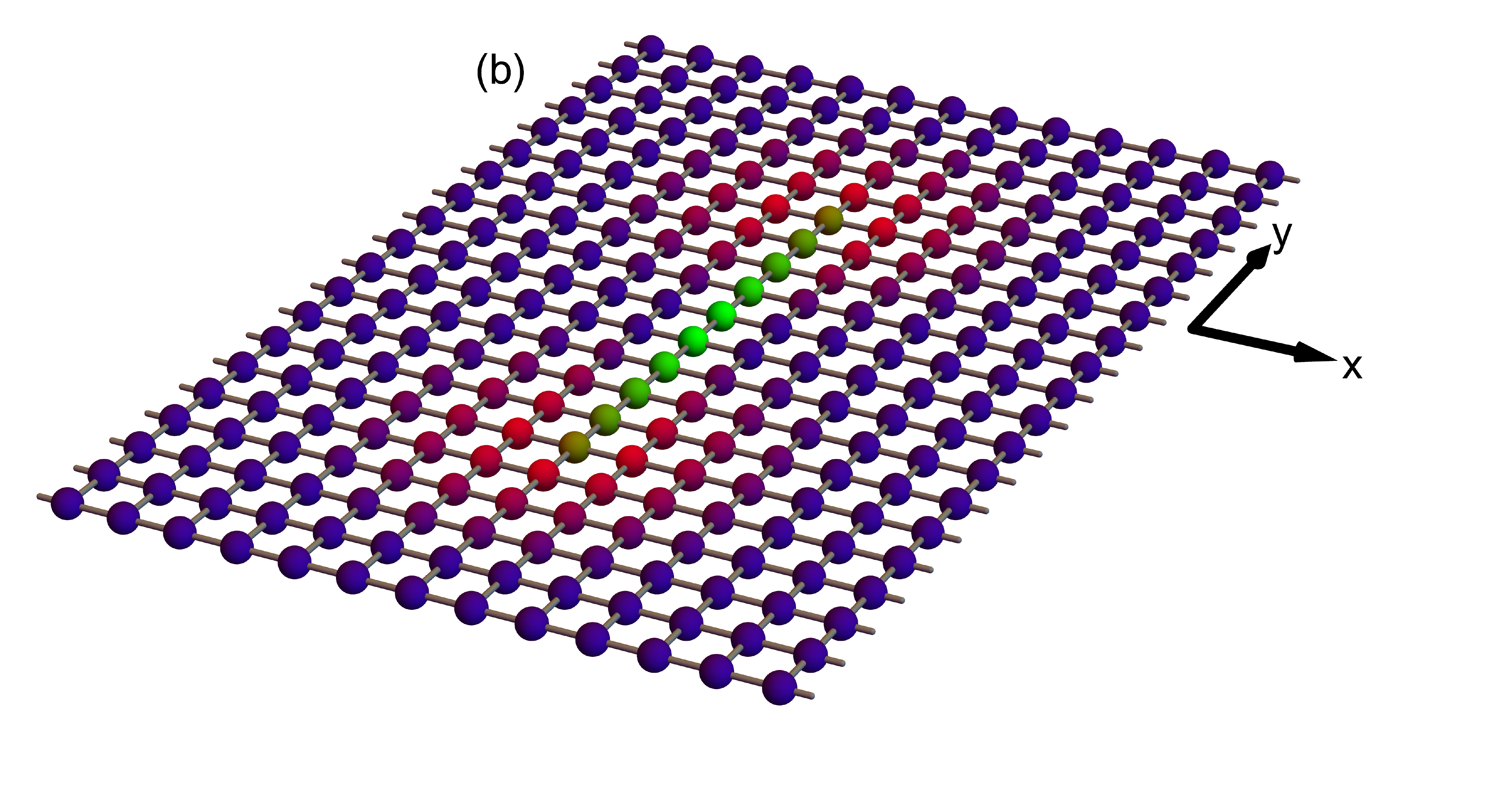}
    \caption{Schematics for the quasi-1D invariant setup (a) versus the numerical TB setup (b). Green sites show the locations of the impurities and a possible localisation of the MZMs is shown in red.}
\label{fig0}
\end{figure}

Here we revisit this analysis, as well as we complete it by considering chains of magnetic impurities besides the scalar  impurity chains. We perform both tight-binding numerical calculations of the local density of states (LDOS) and Majorana polarization of the lowest energy states \cite{Sticlet2012a,Sedlmayr2015b,Sedlmayr2016,Glodzik2020}, as well as calculations of the topological invariant of the chain based on its effective Green's functions \cite{Pinon2020,Sedlmayr2021a}, and of the quasi-one-dimensional topological invariant \cite{Sedlmayr2016,Sedlmayr2021a}, the latter two for the magnetic impurity case only. We found that when the substrate itself is topological the Majorana states forming in the chains are not fully localized, they leak in the bulk and hybridize, for both the magnetic and scalar impurity chains. On the other hand, the magnetic impurity chain may sustain Majorana states for a broad range of parameters, that are fully localized in the wire when the substrate is non-topological and as long as a small p-wave SC component is present. This may occur also for scalar chains but in a much more finely-tuned parameter regime close to the topological transition of the substrate. We analyze also configurations of scalar and magnetic impurity chains on a pure p-wave substrate and we conclude that a pure p-wave SC is not a good candidate to support topological Shiba chains.

The paper is organized as follows: In section II we present the substrate and impurity chain model and the techniques that we will be using. In Section III we present the results for the magnetic and scalar impurity chains. We conclude in Section IV.

\section{Models and techniques}
\label{models}

\subsection{Model}
We consider a two dimensional (2D) lattice with a line of embedded impurities. The corresponding momentum-space Hamiltonian for the superconducting substrate in the basis $\psi_{\bm k}=(c_{{\bm k},\uparrow},c_{{\bm k},\downarrow},c_{-{\bm k},\downarrow}^\dagger,-c_{-{\bm k},\uparrow}^\dagger)^T$ can be written as:
\begin{eqnarray}
	\mathcal{H}^{\rm 2D}_{\bf k}&=&-\left[\mu+2t\left(\cos k_x+\cos k_y\right)\right]\btau^z\otimes\mathbb{I}_2\\\nonumber&&
	-\Delta\btau^x\otimes\mathbb{I}_2 
	+2\kappa \btau^x\otimes\left[\sin k_x\bsig^y-\sin k_y \bsig^x \right]  \,,
\end{eqnarray}
with $\mu$ being the chemical potential, $t$ the nearest-neighbour hopping strength, $\Delta$  and $\kappa$ the s-wave and p-wave superconducting components respectively (both assumed to be real), $\tau$ and $\sigma$ the Pauli matrices in the electron/hole and spin space respectively, and $\mathbb{I}_2 $ a $2 \times 2$ identity matrix.  The impurities are modeled as on-site potentials ${\bm V}=J\btau^z\otimes\bsig^z $ (magnetic impurities) or ${\bm V}=U\btau^z\otimes\mathbb{I}_2 $ (scalar impurities), and the real-space Shiba chain Hamiltonian can be written as: 
\begin{equation}
	\mathcal{H}^{\mathrm{imp}} = \sum\limits_{\bs{r} \in C}\Psi^\dagger_{\bs{r}}{\bm V}\Psi_{\bs{r}}\,,
	\label{eq:HChain}
\end{equation}
with $C$ describing the sites of the one-dimensional chain. 

\subsection{Techniques}

In order to analyze the topological character of the Shiba impurity chains we will use some standard techniques such as numerical tight-binding as well as techniques that we have introduced in a recent work \cite{Sedlmayr2021a}. 

\subsubsection{Effective topological invariant for the Shiba chain}\label{topi}

Along the lines of \cite{Sedlmayr2021a} we calculate a topological invariant for the chain based on its 1D effective Green's function $\mathcal{G}$. Thus, for the Hamiltonian $\mathcal{H}^{\rm 2D}_{\bm k}$ we start with defining the Green's function $\mathcal{G}_0(\omega_n,{\bf k})=(\im\omega_n+\mathcal{H}^{\rm 2D}_{\bm k})^{-1}$. The impurity chain is modeled by adding an infinite line described by the potential ${\bf V}\delta(x)$. The resulting system can be solved by calculating the $T$-matrix \cite{Byers1993,Salkola1996,Ziegler1996,Mahan2000,Balatsky2006,Bena2016}:
\begin{equation}
{\bf T}=\left[\mathbb{I}_4-{\bf V}\mathcal{G}_1(\omega,k_y)\right]^{-1}{\bf V}\,,
\end{equation}
where
\begin{equation}
\mathcal{G}_1(\omega,k_y)\equiv\mathcal{G}_1(\omega,x=0, k_y) =\int_{-\pi}^{\pi} \frac{dk_x}{2\pi}\mathcal{G}_0(\omega,{\bs{k}})\,.
\label{eq5}
\end{equation}
The effective Green's function of the Shiba chain ($x=0$) is obtained from the T-matrix formalism:
\begin{equation}
	\mathcal{G}(\omega,k_y)=\mathcal{G}_1(\omega,k_y)+\mathcal{G}_1(\omega,k_y){\bf T}\mathcal{G}_1(\omega,k_y)\,.
	\label{eq:EffGF}
\end{equation}
We set $k_y\to k$ where possible. This Green's function defines an effective Hamiltonian for the chain: $\mathcal{H}^{\rm 1D}_k\equiv\mathcal{G}^{-1}(0,k)$. 

The model under consideration can have several symmetries important for the topology~\cite{Chiu2016,Sato2017}:
\begin{enumerate}[(i)]
%%%%%%%%%
\item Particle--hole (PH) symmetry described by an anti-unitary operator $\mathcal{P}=\btau^y\otimes\bsig^y\mathcal{K}$, such that $\left[\mathcal{P},\mathcal{H}_{k}\right]_+=0$ and $\mathcal{P}^{2} = 1$. $\mathcal{K}$ is the complex conjugation operator.
%%%%%%%%%
\item The ``time-reversal'' (TR) symmetries described by the anti-unitary operators $\mathcal{T}_+=\im\btau^0\otimes\bsig^z\mathcal{K}$ and $\mathcal{T}_-=\im\btau^0\otimes\bsig^y\mathcal{K}$, where $\left[\mathcal{T}_{\pm},\mathcal{H}_{k}\right]_-=0$ with $\mathcal{T}_\pm^{2} = \pm1$.
%%%%%%%%%
\item Finally we have the combination of PH and TR symmetries, the sublattice or ``chiral'' symmetry described by the unitary operators $\mathcal{S}_+=\btau^y\otimes\bsig^x$ and $\mathcal{S}_-=\im\btau^y\otimes\bsig^0$, with $\left[\mathcal{S}_\pm, \mathcal{H}_{k}\right]_-=0$.
%%%%%%%%%
\end{enumerate}
The substrate, in the absence of any impurity potential, has both $\mathcal{P}$ and $\mathcal{T}_-$ symmetries, placing it in the DIII class with Kramer's pairs of topologically protected modes in its topologically non-trivial phases~\cite{Ryu2010}. The $\mathcal{T}_+$ symmetry is broken by the p-wave term in the $x$-direction. The addition of scalar impurities does not change the symmetries, however adding magnetic impurities breaks $\mathcal{T}_-$. The quasi-1D set-up with magnetic impurities therefore possesses only particle-hole symmetry and is in class D with a $\mathbb{Z}_2$ invariant. In the strict 1D limit of the chain the $\mathcal{T}_-$ symmetry is recovered, and therefore for the magnetic chain we have both $\mathcal{P}$ and $\mathcal{T}_+$ symmetries, giving BDI in the classification which has a chiral $\mathbb{Z}$ invariant. For the chain of electronic impurities one must be more careful. In this case we have $\mathcal{P}$, $\mathcal{T}_-$, and $\mathcal{T}_+$. Therefore one should first diagonalise with respect to the unitary symmetry $\mathcal{U}=\mathcal{T}_-\mathcal{T}_+$. In each diagonalised subblock only a chiral symmetry will remain and the system is in class AIII~\cite{Ryu2010}.

For the magnetic impurity chain we can therefore use a standard 1D chiral invariant \cite{Gurarie2011,Sedlmayr2021a}
\begin{equation}\label{chi_inv}
\nu_{\rm Ch}=\frac{1}{4\pi \im}\int_{-\pi}^{\pi}dk\Tr\mathcal{S}_+\mathcal{H}^{\rm 1D}_k\partial_k\left[\mathcal{H}^{\rm 1D}_k\right]^{-1}\,.
\end{equation}
If there are no magnetic impurities then we still have standard time reversal symmetry, and this method no longer captures the appropriate invariant.

\subsubsection{Numerical tight-binding analysis}

We perform a numerical exact diagonalization of the real-space tight-binding model in a periodic-boundary-conditions setup to calculate the wave function corresponding to the lowest-energy state. Choosing periodic boundary conditions ensures that the edge states associated with the boundaries of the substrate do not interfere with the states localized on the wire. Using the techniques derived in \cite{Sedlmayr2015b} and \cite{Sedlmayr2016} we calculate the Majorana polarization associated with the lowest energy states and we plot it as a function of position~\cite{Sedlmayr2015b,Sedlmayr2016,Glodzik2020,Sedlmayr2021a,Maska2021}. In most of our analyses we fix the s-wave SC component and we modify the p-wave component. The corresponding topological chain phase diagram is derived by  calculating the total Majorana polarization summed on half the wire~\cite{Sedlmayr2015b} as a a function of the SC p-wave parameter and the impurity strength (scalar or magnetic). We also calculate the local Majorana polarization vector and we plot it as a function of position for particular parameter sets. Finally we also use this technique to derive the topological phase diagram in the space of the chemical potential value and impurity strength.

The local Majorana polarization for a state $|\psi_n\rangle$, with a wavefunction $(u_{{\bf r},\uparrow},u_{{\bf r},\downarrow},v_{{\bf r},\downarrow},v_{{\bf r},\uparrow})$ at position $\bf r$, is defined as
\begin{equation}
 M_{\bf r}= \e^{\im\varphi}\langle\psi_n|\hat{\bf r}\mathcal{P}|\psi_n\rangle=2\e^{\im\varphi}\left(u^*_{{\bf r},\downarrow}v^*_{{\bf r},\downarrow}-u^*_{{\bf r},\uparrow}v^*_{{\bf r},\uparrow}\right)\,,
\end{equation}
where $\hat{\bf r} \mathcal{P}$ is the spatial projection of the particle-hole operator $\mathcal{P}$. The overall phase $\varphi$ is arbitrary. The total Majorana polarization is therefore
\begin{equation}
 M=\left|\sum_{\bf r\in C'}M_{\bf r}\right|\,,
\end{equation}
with $C'$ being the set of all the sites in the left half of the impurity chain. For a well defined MZM one has
\begin{equation}
	M=\rho'\equiv\sum_{\bf r\in C'}\rho_{\bf r}\,,
\end{equation}
with
\begin{equation}
 \rho_{\bf r}=\langle\psi_n|\hat{\bf r}|\psi_n\rangle=\sum_\sigma\left[\left|u_{{\bf r},\sigma}\right|^2+\left|v_{{\bf r},\sigma}\right|^2\right]
\end{equation}
denoting the total density of the state $|\psi_n\rangle$ on site $\bf r$.

When time reversal symmetry is present, see section \ref{topi}, we must also be careful with the states that we consider in the degenerate subspace. In this case the MZMs always come in degenerate Kramer's pairs, with one pair localized at one end of the chain. If we consider two such degenerate MZMs $|\psi_{01}\rangle$ and $|\psi_{02}\rangle$, it is clear that an arbitrary combination of the two is no longer an eigenstate of the particle-hole operator. For example, if $\mathcal{P}|\psi_{01,2}\rangle=|\psi_{01,2}\rangle$ then
\begin{multline}
	\mathcal{P}\left[a\e^{\im\alpha}|\psi_{01}\rangle+b\e^{\im\beta}|\psi_{02}\rangle\right]=a\e^{-\im\alpha}|\psi_{01}\rangle+b\e^{-\im\beta}|\psi_{02}\rangle\\
	\neq\e^{\im\gamma}\left[a\e^{\im\alpha}|\psi_{01}\rangle+b\e^{\im\beta}|\psi_{02}\rangle\right]\,,
\end{multline}
where $a,b,\alpha,\beta$, and $\gamma$ are real with $a^2+b^2=1$. For the numerical calculation of the Majorana polarization it is therefore always important to rotate within the degenerate subspace to a basis of two MZMs.

\subsubsection{Invariant from Quasi-1D Strip}
An alternative route to calculating the invariant is to treat the impurity chain plus substrate as a quasi-1D system, consisting of one line of impurity sites and on either side a number of lines of sites of the background 2D lattice, totalling $N_x$ lines \cite{Sedlmayr2016,Sedlmayr2021a}. We then calculate the invariant for different $N_x$. In the limit that $N_x$ is large the invariant should no longer depend explicitly on $N_x$. 
The invariant is given by the parity of the negative energy bands at the time reversal invariant momenta~\cite{Sato2009b}. Following an appropriate rotation the Hamiltonian can be written at the time reversal invariant momenta as $H^{\rm Q1D}_{0,\pi}\to \textrm{diag} (\bar{\mathcal{H}}(0,\pi),-\bar{\mathcal{H}}(0,\pi))$, and calculating the parity of the negative energy bands is then equivalent to calculating
\begin{equation}%\label{invariant}
\delta=(-1)^\nu_{\rm Q1D}=\sgn\left[\det\bar{\mathcal{H}}(0)\det\bar{\mathcal{H}}(\pi)\right]\,.
\end{equation}
Note that this system does not have chiral symmetry, so it is in the D class which has a $\mathbb{Z}_2$ invariant, and hence we can write
\begin{equation}\label{invariant}
\nu_{\rm Q1D}=\frac{1-\sgn\left[\det\bar{\mathcal{H}}(0)\det\bar{\mathcal{H}}(\pi)\right]}{2}\,.
\end{equation}

We should note that the result of this calculation is not exactly what we are looking for in the present configuration: the quasi-1D topological invariant predicts the formation of chain end states for a setup in which the impurity chain ends exactly at the same position as the strip (see Fig.~\ref{fig0}). This does not capture the same physics as that of a chain fully embedded in an infinite substrate, thus the results obtained using this method are only indicative of the underlying physics, but do not capture for example accurately the destruction of the Majorana modes in the chain via a leakage in the substrate. We will thus use them as a basis and comparison for the rest of the results. Also note once more that this technique only works for magnetic impurities and not for scalar ones.

\section{Results}

\subsection{Magnetic impurities}

\begin{figure}
  \includegraphics[height=0.4\columnwidth]{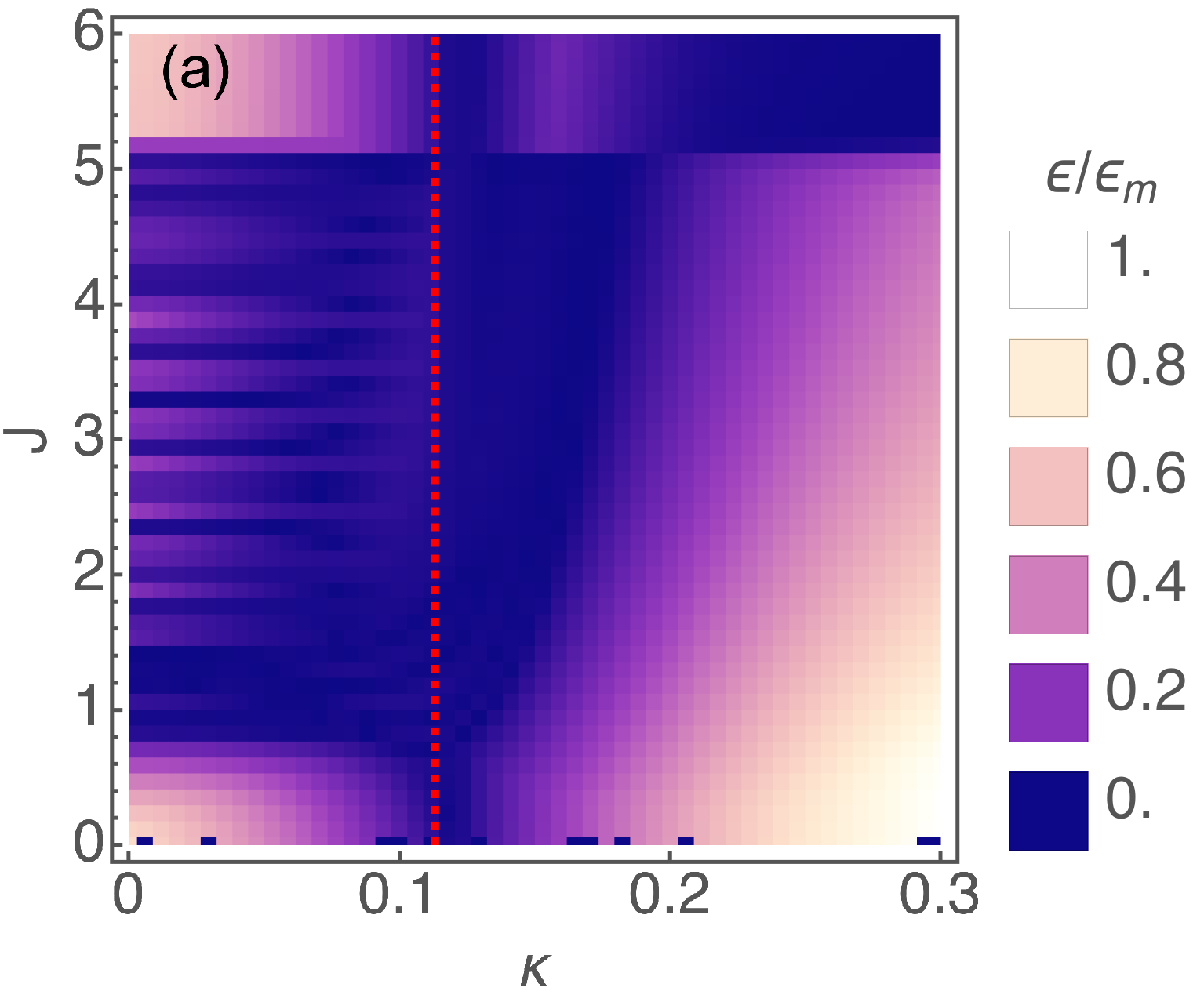}
  \includegraphics[height=0.4\columnwidth]{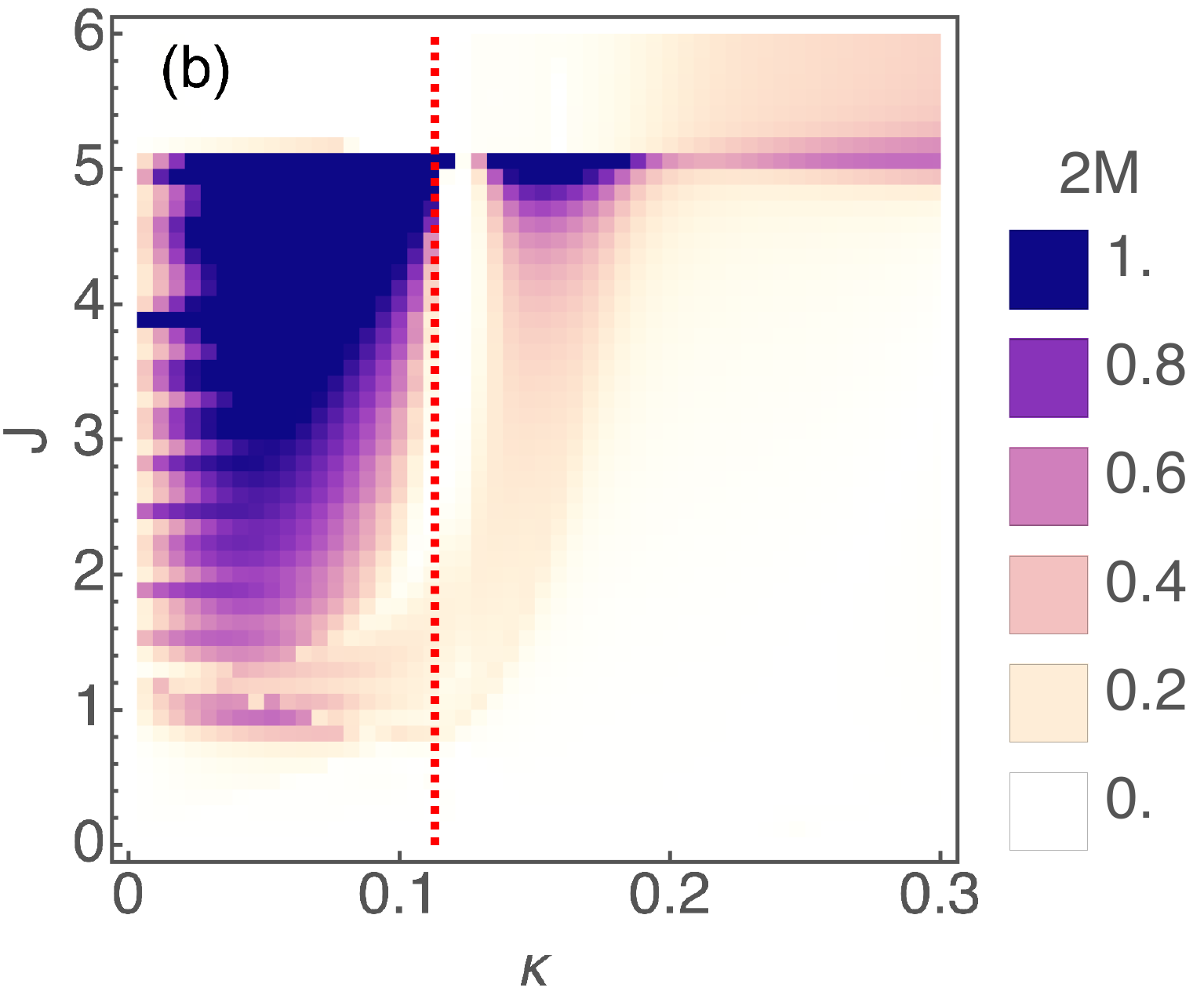}\\
  \includegraphics[height=0.4\columnwidth]{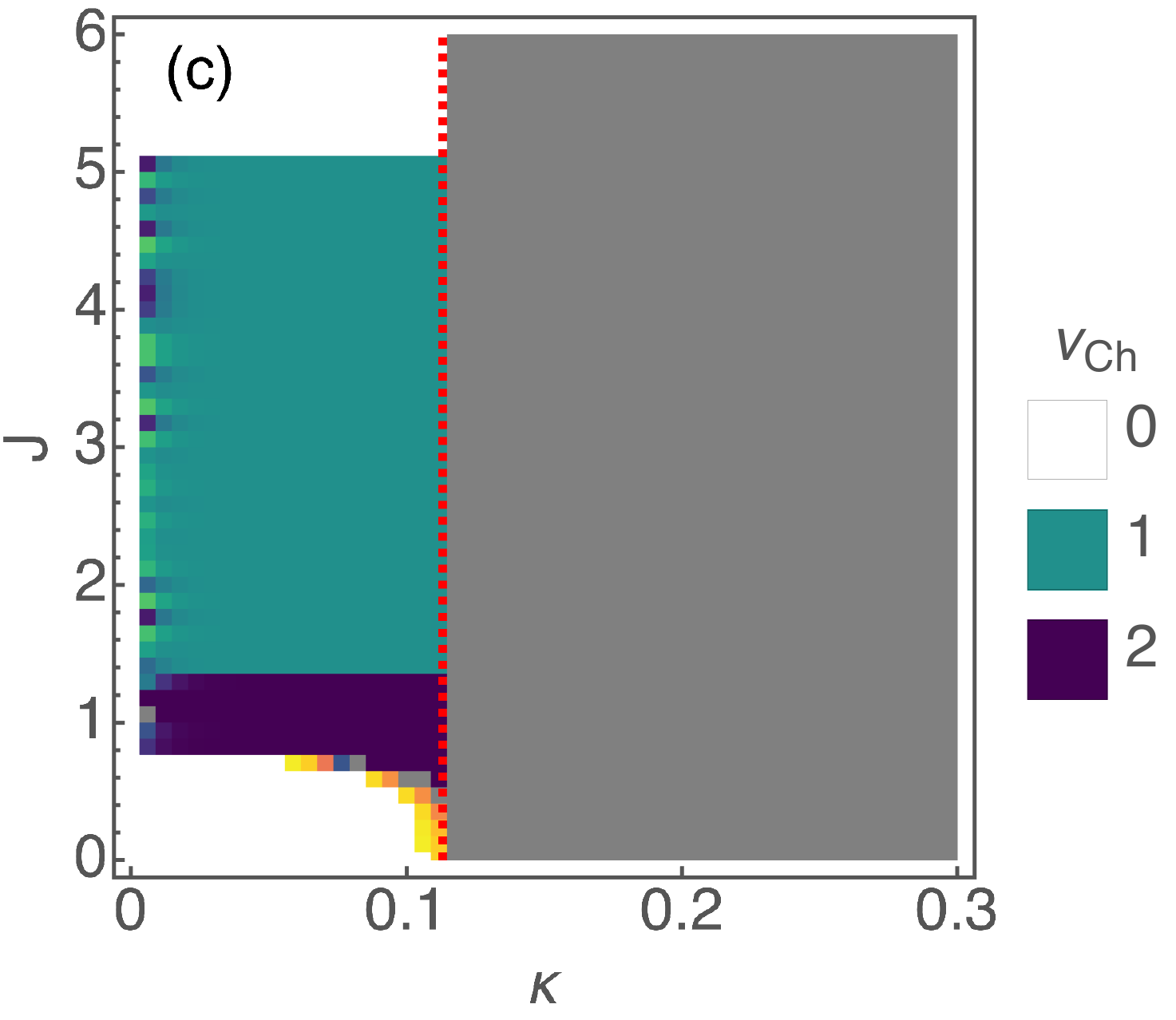}\hspace{0.3cm}
  \includegraphics[height=0.4\columnwidth]{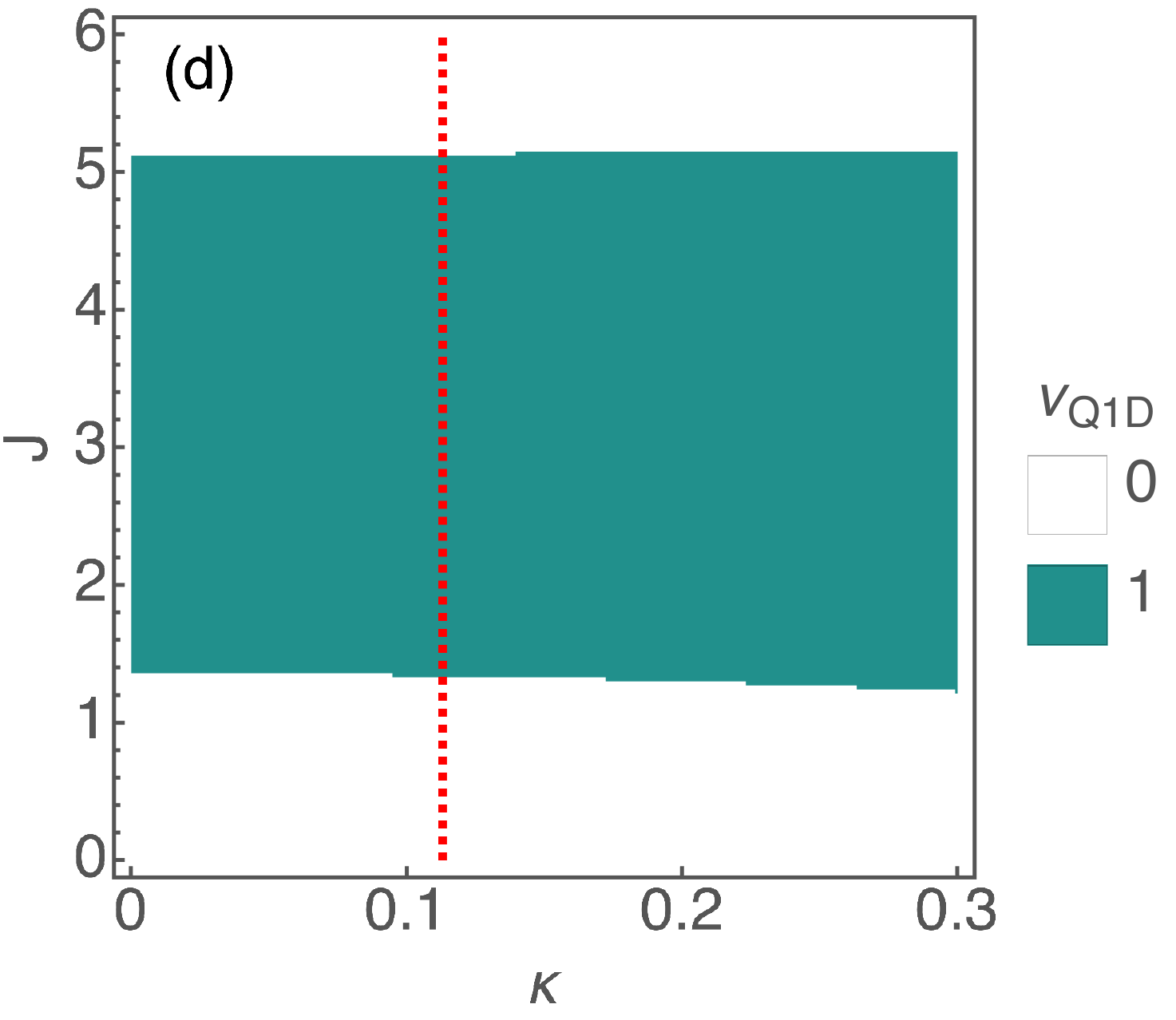}
  \caption{ A comparison of the lowest energy state energy(a), Majorana polarization (b), GF invariant (c) and quasi-1D invariant (d) for a chain of magnetic  impurities within a 2D $s$-wave/$p$-wave lattice, as a function of the p-wave parameter $\kappa$ and the magnetic impurity strength $J$. Here $\Delta= 0.16t$ and $\mu=3.5$. The system size is $21\times 80$ with a gap of $20$ sites between the boundary of the system and the end of the chain, periodic boundary conditions are imposed.  The red dashed line is the topological phase boundary $2\kappa^2(4-\mu)=\Delta^2$ of the substrate. The grey region in (c), which matches the region where the substrate is topologically non-trivial, is where the integral for the invariant did not converge in time. The maximum energy in the phase diagram ($\epsilon_m=0.225t$) is used to scale the energy plot.}
\label{magimpmupd}
\end{figure}

In Fig.~\ref{magimpmupd} we plot the energy and the Majorana polarization of the lowest energy state, as well as the GF invariants as a function of the p-wave SC component and the magnetic impurity strength. The s-wave SC component is fixed at $\Delta=0.16$, same as in Ref.~\onlinecite{Neupert2016}.

\begin{figure}
 	\includegraphics[width=0.9\columnwidth]{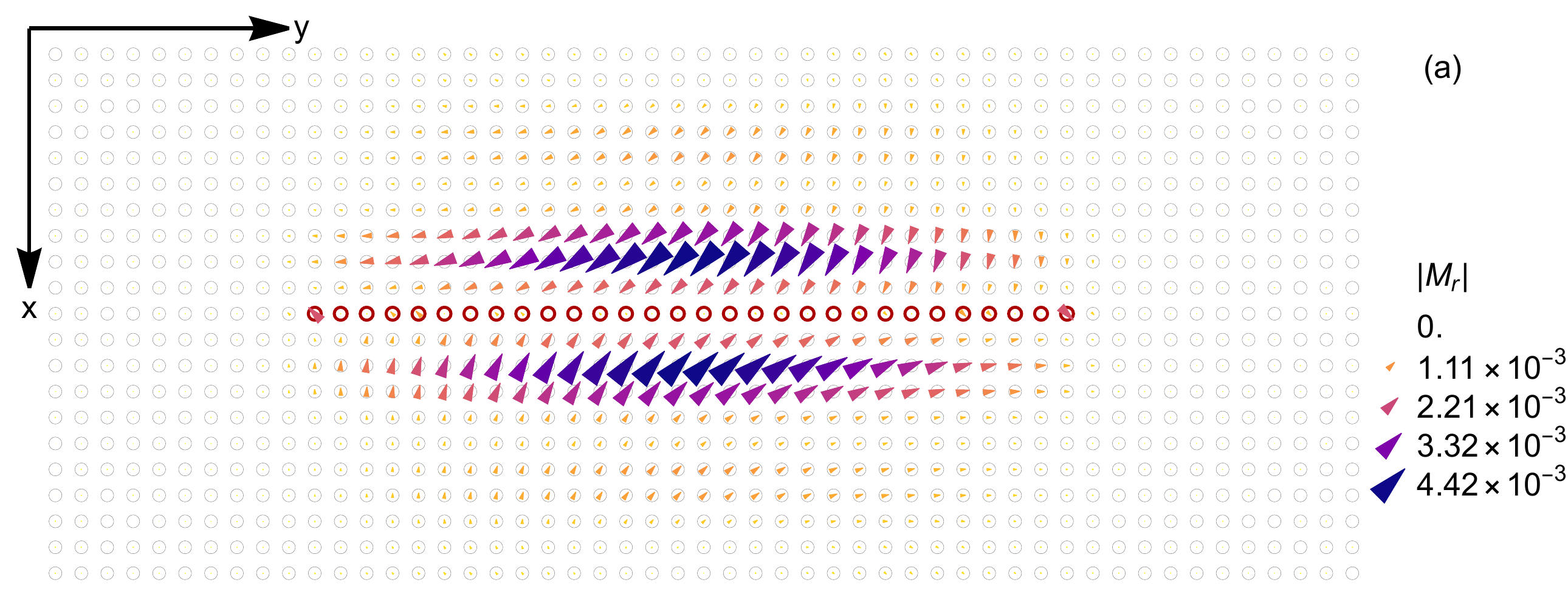}\\
 	\includegraphics[width=0.9\columnwidth]{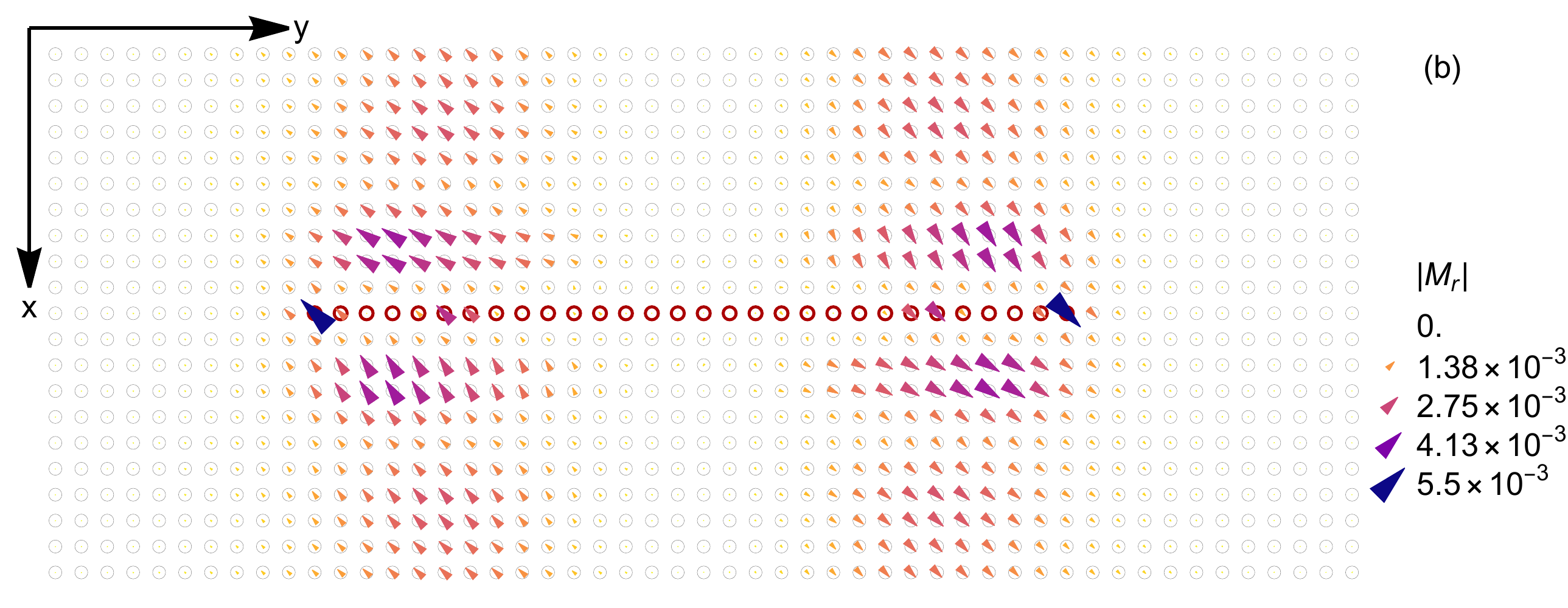}
  \caption{The Majorana polarization for two points in the phase space for which the substrate is topological. We take $\Delta= 0.16t$ and $\mu=3.5t$. For panel (a) $J= 3t$, $\kappa=0.2t$, and for panel (b) $J= 1.8t$, $\kappa=0.14t$. Note that the Majorana polarization vector precesses and that the lowest energy states leak into the bulk either parallel to the wire or close to the ends, see panels (a) and (b) respectively). The system size is $21\times 70$ with a gap of $20$ sites between the boundary of the system and the end of the chain, periodic boundary conditions are imposed.
}\label{mpm1}
\end{figure}

Note that the quasi-1D invariant predicts a wide range of parameters for which the system is topological. As described in Fig.~\ref{fig0} this corresponds to cutting the entire system in half and recovering end states when the end of the chain corresponds exactly to the end of the wire. However, this is not what happens in more realistic setups, thus the phase diagram obtained by calculating the Majorana polarization in the wire is more relevant for an experimental situation corresponding to that in Fig.~\ref{fig0}b). The corresponding phase diagram indicates that the Majorana states forming in the wire leak significantly in the bulk if the substrate is topologically non-trivial (for $\kappa>\Delta/2\sqrt{4-\mu}$). In this regime only a part of the weight of the lowest-energy states is localized in the wire, while the rest is distributed in the bulk, either in the vicinity of the end, or parallel to the wire itself. Such configurations are shown in Figs.~\ref{mpm1} (a) and (b). We note that, moreover, the local Majorana polarization vectors in this situation are not aligned inside a certain region (that would be consistent to the formation of a proper Majorana state), but precess, indicating that the corresponding states are not actual Majorana, but quasi-Majorana, even if their energy is very close to zero. They thus cannot be fully separated and could not be properly used as qubits.

\begin{figure}
 	\includegraphics[width=0.9\columnwidth]{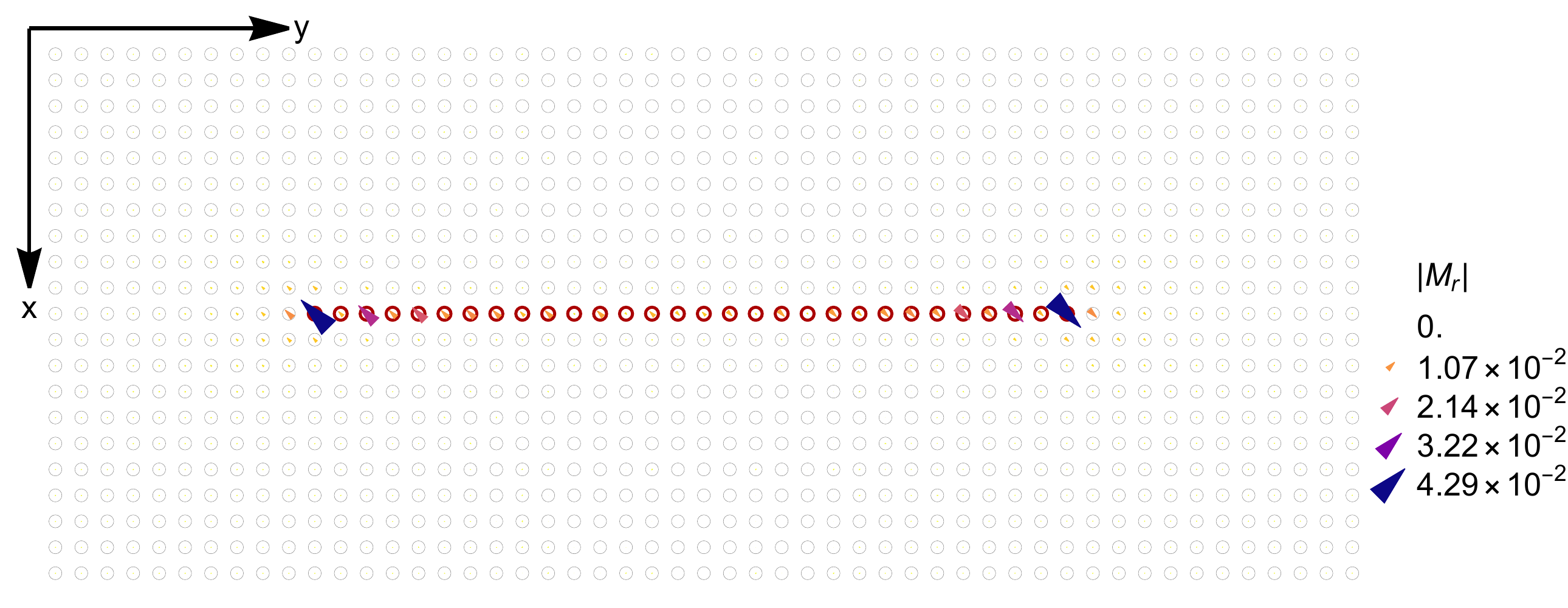}
  \caption{The Majorana polarization for a point in the phase space for which the substrate is non-topological. We take $J= 3t$, $\Delta= 0.16t$, $\mu=3.5t$, and $\kappa=0.05t$. Note that the Majorana polarization vector of the lowest energy states is concentrated mainly on the chain. The system size is $61\times 70$ with a gap of $20$ sites  between the boundary of the system and the end of the chain, periodic boundary conditions are imposed.
}\label{mpm2}
\end{figure}

When the substrate is topologically trivial (for $\kappa<\Delta/2\sqrt{4-\mu}$) the lowest-energy states actually become fully localized in the chain and the Majorana polarization is getting close to 1, indicating very small leaks in the bulk. In Fig.~\ref{mpm2} we show the local Majorana polarisation in such a situation.

Note that when the substrate is trivial, the results for the topological character of the wire obtained by calculating the GF invariant (Fig.~\ref{magimpmupd}c) are similar to those obtained by the Majorana polarization calculation (Fig.~\ref{magimpmupd}b) and the quasi-1D  invariant  (Fig.~\ref{magimpmupd}d). However, when the substrate is topologically non-trivial the integral for the GF invariant in Eq.~\eqref{chi_inv} converges very slowly: due to the complex form of the integrand the integral over $k$ is implemented as a simple Riemann sum with a number of steps $N_{\rm Ch}=200$. For the topologically non-trivial substrate this integral does not converge with $N_{\rm Ch}$, and a scaling analysis shows that not only will it not converge except for numerically unattainable $N_{\rm Ch}$, but also that in the limit of $N_{\rm Ch}\to\infty$ we will have $\nu_{\rm Ch}\to1$ irrespective of whether the wire itself is topological or not, indicating that in this particular case the invariant tracks rather the topological character of the 2D substrate.

\begin{figure}
	\includegraphics[height=0.4\columnwidth]{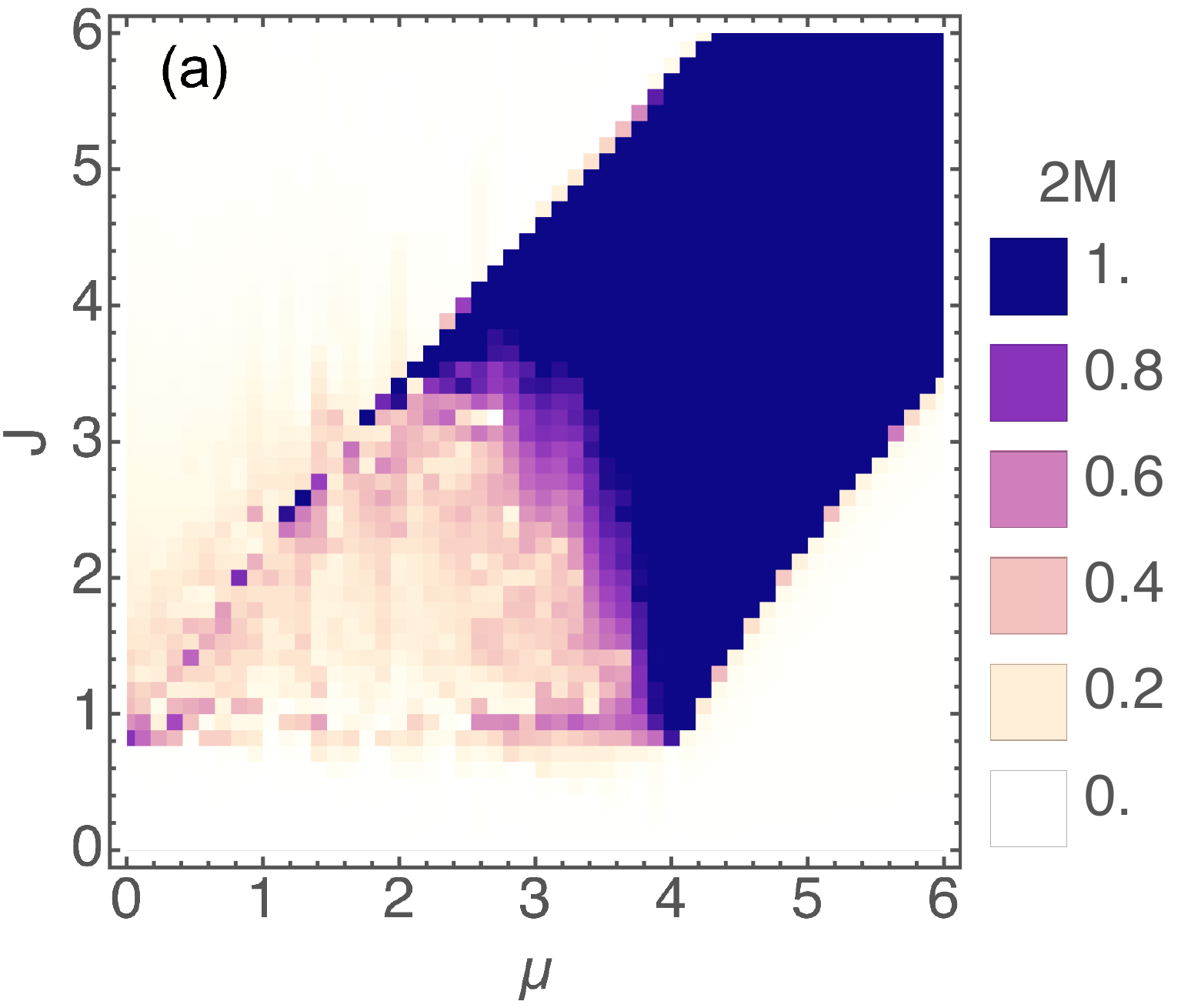}
	\includegraphics[height=0.4\columnwidth]{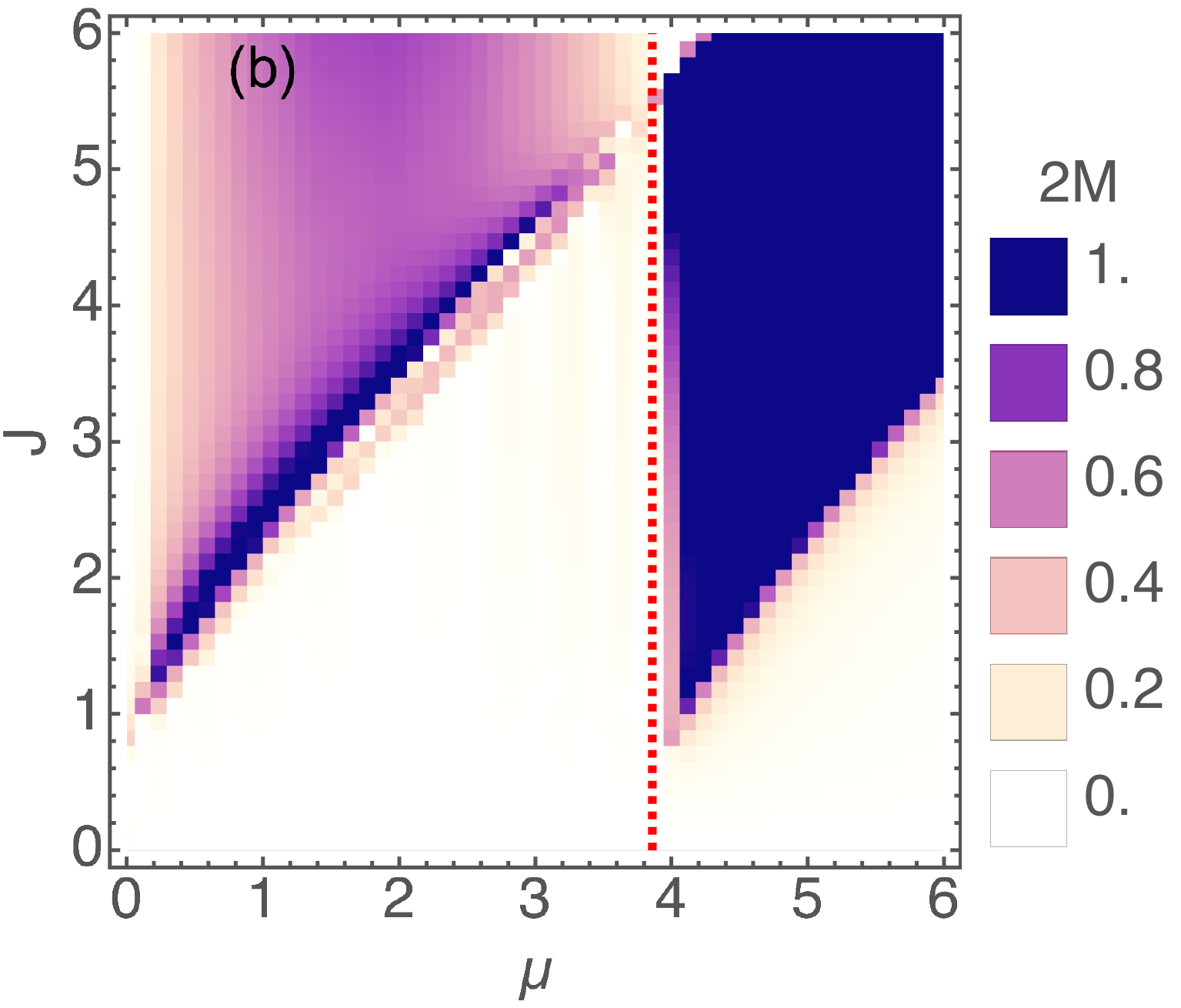}\\
	\includegraphics[height=0.4\columnwidth]{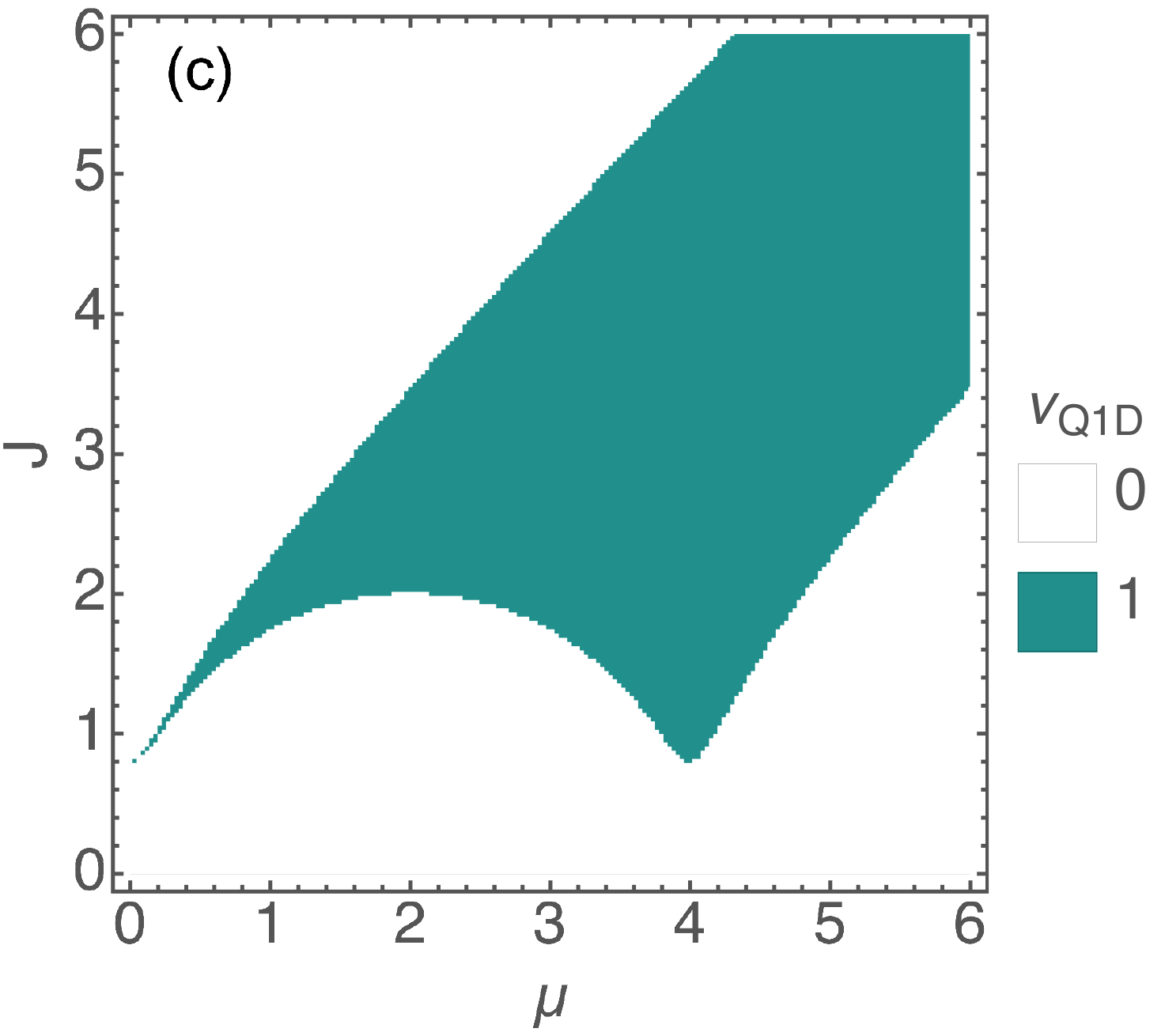}\hspace{0.2cm}
	\includegraphics[height=0.4\columnwidth]{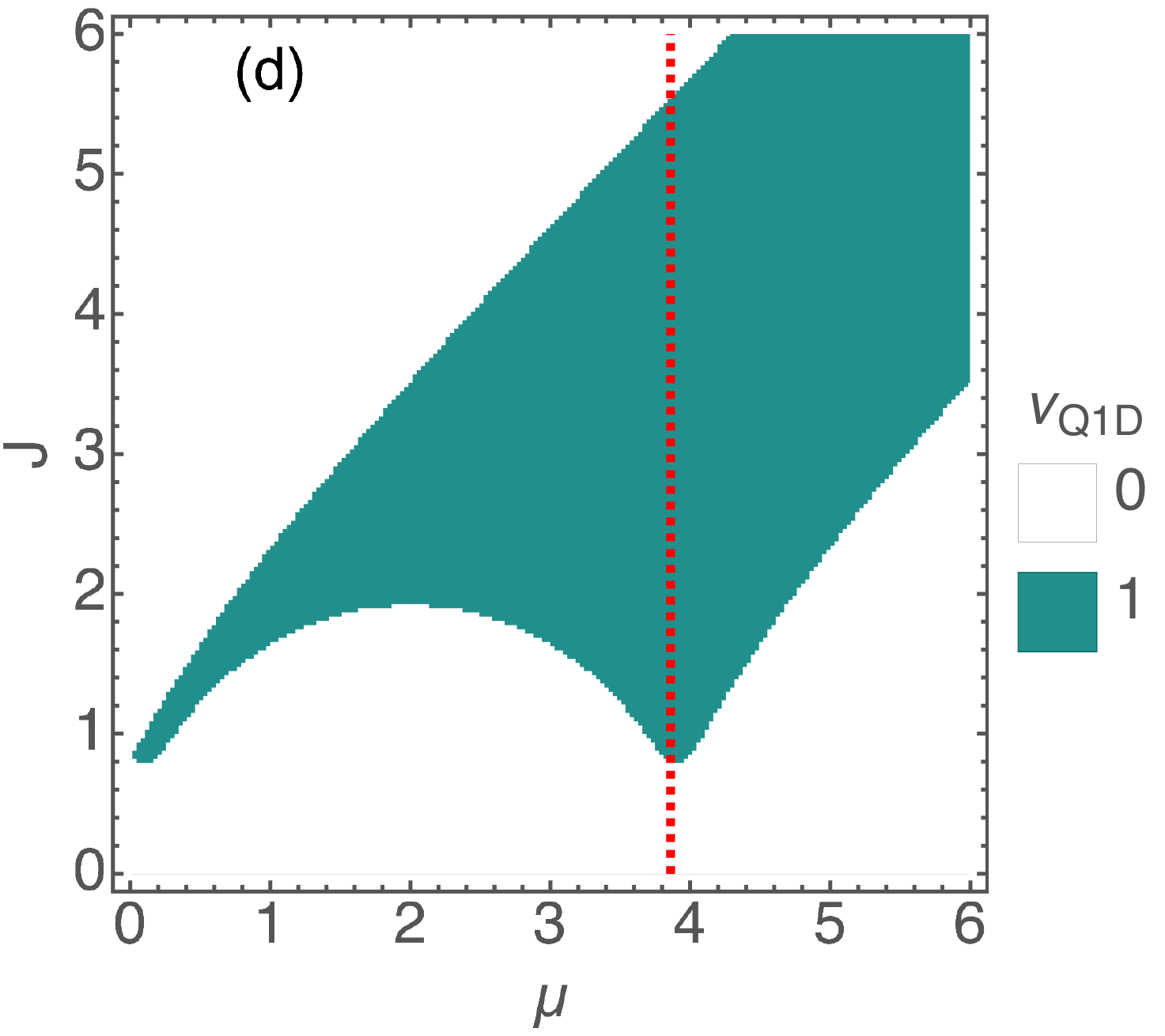}
  \caption{The Majorana polarization and the quasi 1D invariant as a function of $\mu$ and $J$. Here $\Delta= 0.16t$, with  $\kappa=0.05t$ (on the left) and  $\kappa=0.3t$ (on the right). The system size is $21\times 80$ with a gap of $20$ sites between the boundary of the system and the end of the chain, and we impose periodic boundary conditions. The red dashed line is the topological phase boundary $2\kappa^2(4-\mu)=\Delta^2$ of the substrate. For (a,c) there is no non-trivial phase in the substrate.}
  \label{muJ}
\end{figure}

We note that the physics of a magnetic impurity chain on a trivial substrate with a non-zero p-wave component is very similar to that for an s-wave SC substrate with Rashba spin-orbit coupling. It appears that the p-wave component of the SC, no matter how small, would play the role of the spin-orbit coupling, and thus Majorana states would form also for a chain of magnetic impurities deposited on a substrate with a small but non-zero p-wave component. To check this we consider a combination of an s-wave component $\Delta=0.16t$ and a p-wave one $\kappa=0.05t$, and we plot the Majorana polarization as a function of $\mu$ and $J$ (see Fig.~\ref{muJ} top left panel). Indeed we recover exactly the same profile as in the topological phase diagram as function of $\mu$ and $J$ in the presence of spin-orbit coupling in the substrate~\cite{Sedlmayr2021a}. This makes us believe that the Rashba spin-orbit coupling is equivalent to a small p-wave component in what concerns the formation of Majorana states in chains of magnetic impurities. We note that for a larger p-wave component, $\kappa=0.3t$ (see Fig.~\ref{muJ}  top right panel) the corresponding ($\mu$, $J$) topological phase diagram indicates that the wire becomes topological only when the substrate is not. 

\begin{figure}
 \includegraphics[width=0.9\columnwidth]{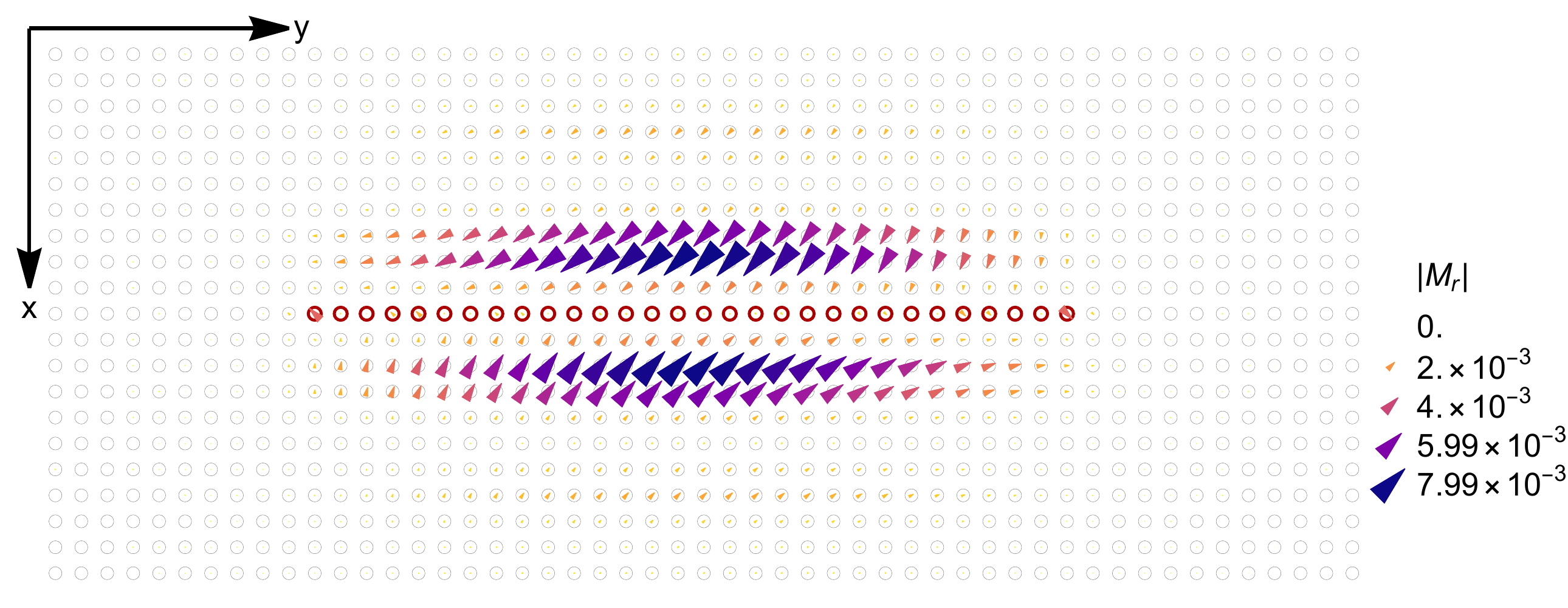}
  \caption{The Majorana polarization for a chain of magnetic impurities on a pure p-wave substrate.  We take $J= 3t$, $\Delta= 0t$, $\mu=3.5t$, and $\kappa=0.2t$. The system size is $21\times 70$ with a gap of $20$ sites between the boundary of the system and the end of the chain, periodic boundary conditions are imposed.}
  \label{mpm3}
\end{figure}

To conclude this section we consider the case of a pure p-wave, and in Fig.~\ref{mpm3} we plot the Majorana polarization as a function of position for a chain of magnetic impurities. We note that the lowest-energy states in the wire leak strongly in the substrate, consistent with what we observed also for the mixed s+p SC substrate with a dominant p-wave component. Thus we claim that clean and isolated Majorana states cannot be obtained on a topological substrate by depositing magnetic impurities, but they can be obtained when the substrate is non-topological with either a Rashba or a small p-wave component.

\subsection{Scalar impurities}

\begin{figure}
 \includegraphics[height=0.4\columnwidth]{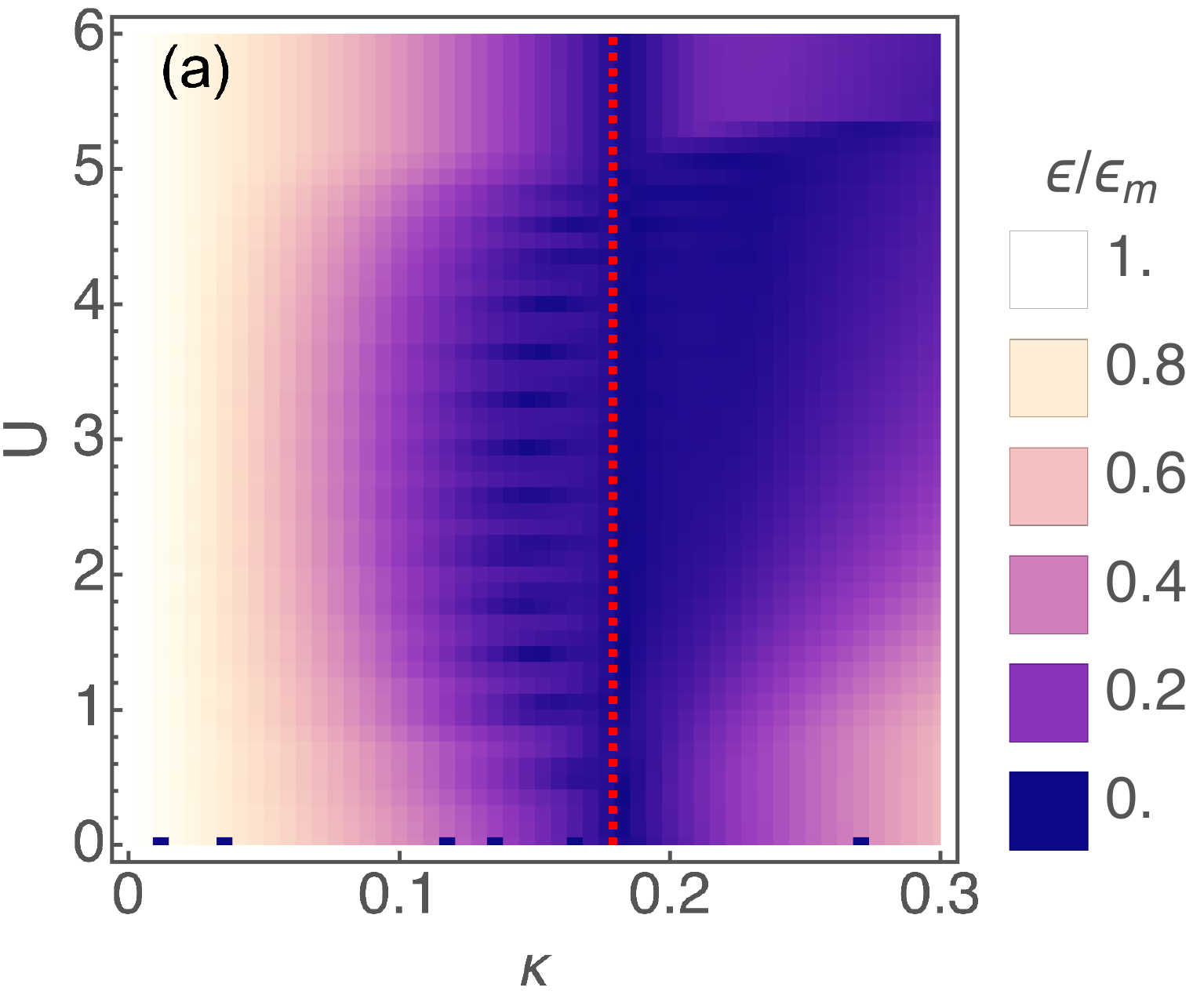}
 \includegraphics[height=0.4\columnwidth]{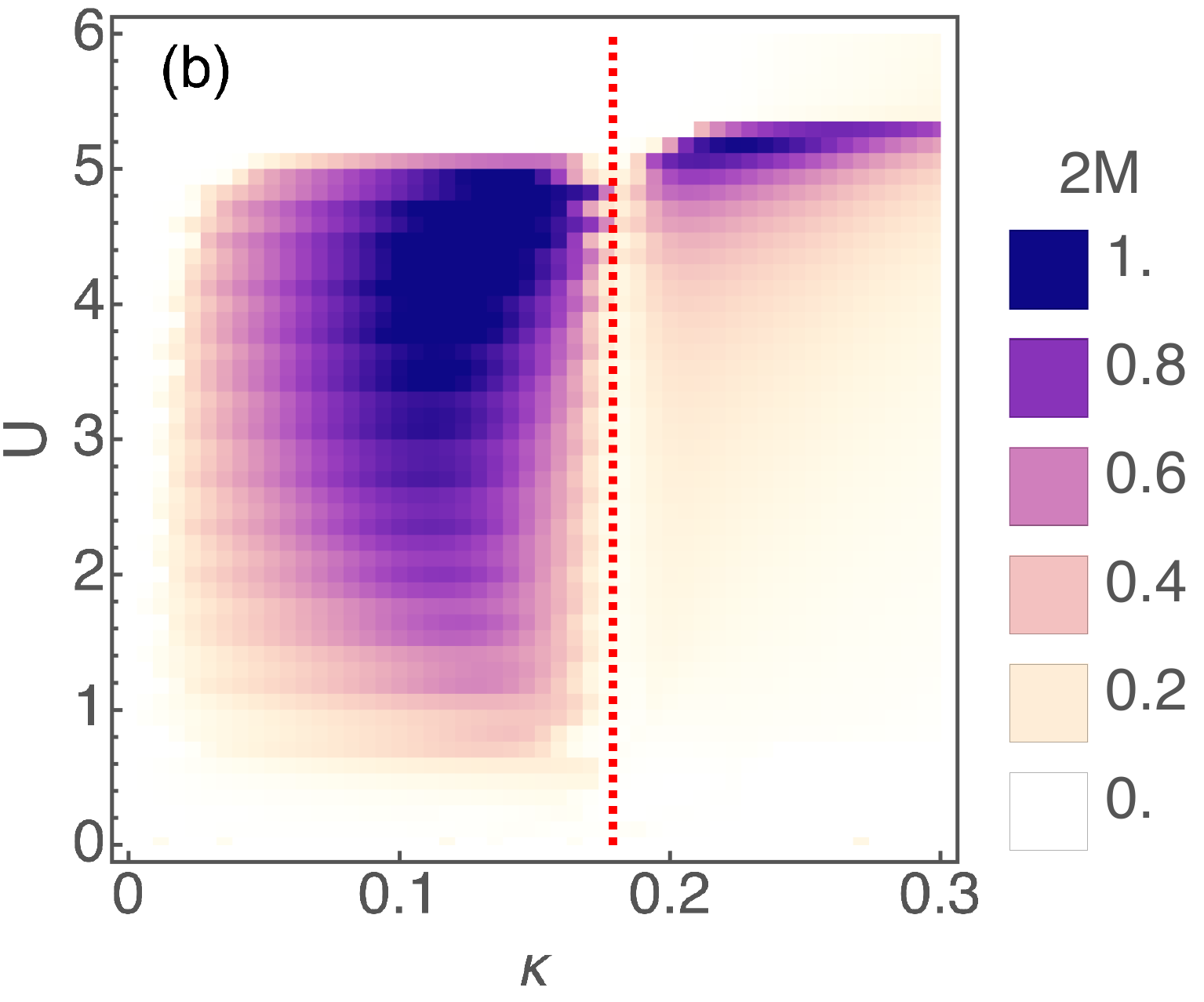}
  \caption{A comparison of the lowest energy state energy and the Majorana polarization for a chain of scalar impurities deposited on a 2D $s$-wave/$p$-wave lattice, as a function of the p-wave parameter $\kappa$ and the scalar impurity strength $U$. Here $\Delta= 0.16t$ and $\mu=3.8t$. The system size is $21\times 80$ with a gap of $20$ sites between the end of the chain and the system boundary, and periodic boundary conditions are imposed.    The red dashed line is the topological phase boundary $2\kappa^2(4-\mu)=\Delta^2$ of the substrate. The maximum energy in the phase diagram ($\epsilon_m=0.225t$) is used to scale the energy plot.}
\label{scalarimpmupd}
\end{figure}

In this configuration the time reversal symmetry is preserved. It is difficult to calculate an appropriate TRS invariant from an effective Green's function, as well as a quasi-1D invariant. Thus we base our analysis on numerical tight-binding and we calculate the topological phase diagrams based on the Majorana polarization of the lowest-energy states. In Fig.~\ref{scalarimpmupd} we plot the Majorana polarization of the lowest-energy state as a function of the the p-wave SC component and the value of the impurity potential U, for a fixed value of the s-wave SC component $\Delta=0.16t$. 

The topological transition between a topological and trivial bulk would take place around $\kappa=\Delta/2\sqrt{4-\mu}$. For larger p-wave pairings the substrate is topological, and for smaller ones it is trivial. From Fig.~\ref{scalarimpmupd}  we can see that the Majorana polarization is significant on the wire only for a very finely tuned region, mostly localized around the bulk topological transition, for a region in which the substrate is non-topological. Moreover, for the system sizes considered, the Majorana polarization values obtained are not perfect as observed for the magnetic impurity chains ($2M \approx 1$), indicating a fragility for these states, whose perfect Majorana character cannot even be fully recovered numerically here. The formation of Majorana states in the presence of a scalar impurity chain on a non-topological substrate was predicted in \cite{Neupert2016}, as well as in \cite{Kaladzhyan2018}. Indeed we recover here the same phenomenon, however, we note that the existence of fully localized and distinct Majorana states is only possible when the substrate is trivial, in a very finely-tuned region, for the rest of the parameter phase space the Majoranas leak in the bulk and they cannot be considered isolated states that would allow for example their use as qubits. We note that Majorana states leaking into adjacent structures is a generic phenomena whenever no energy gap prevents it~\cite{Gibertini2012,Guigou2016,Kobialka2018,Sedlmayr2021}.

\begin{figure}
 \includegraphics[width=0.9\columnwidth]{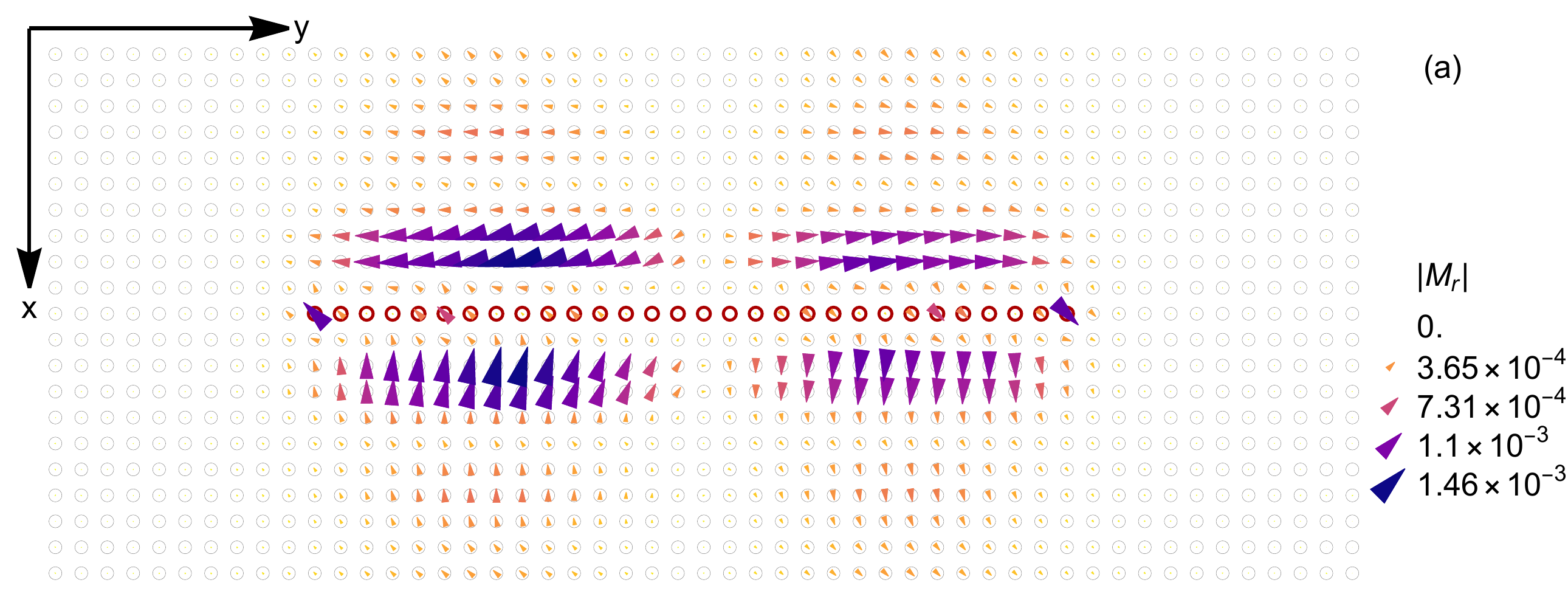}\\
 \includegraphics[width=0.9\columnwidth]{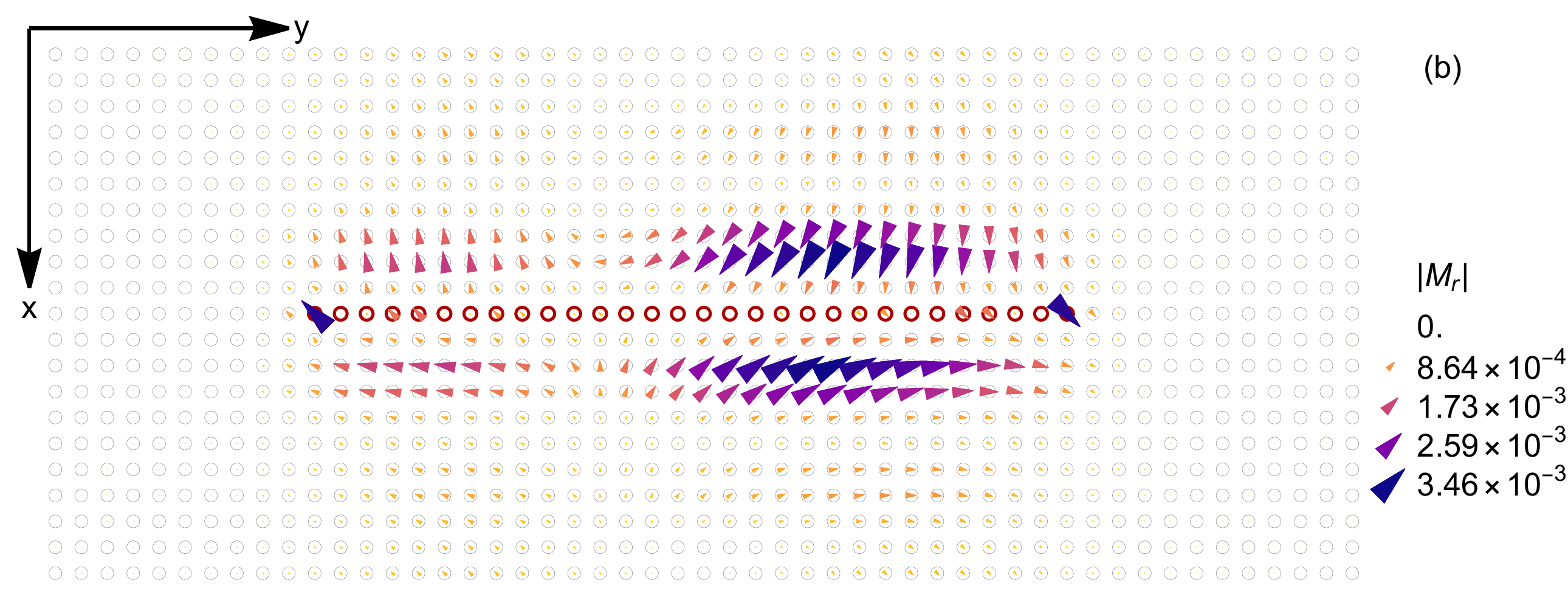}\\
 \includegraphics[width=0.9\columnwidth]{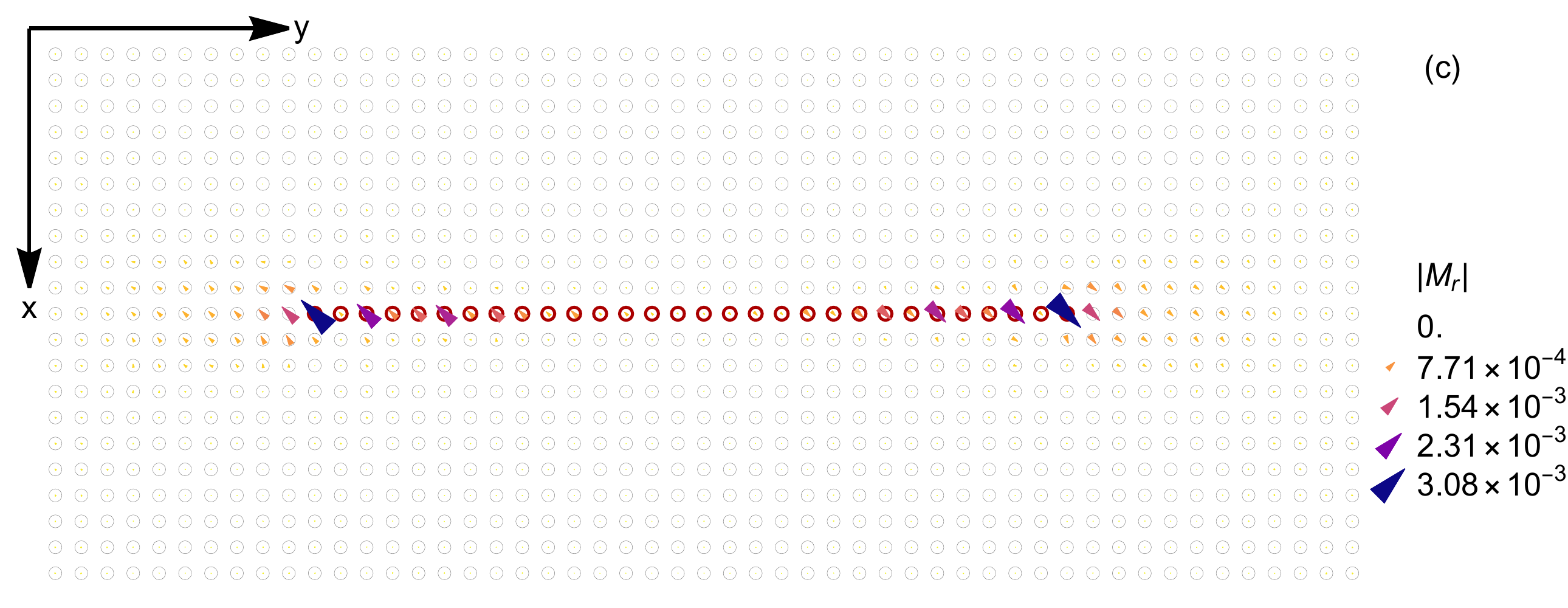}\\
 \includegraphics[width=0.9\columnwidth]{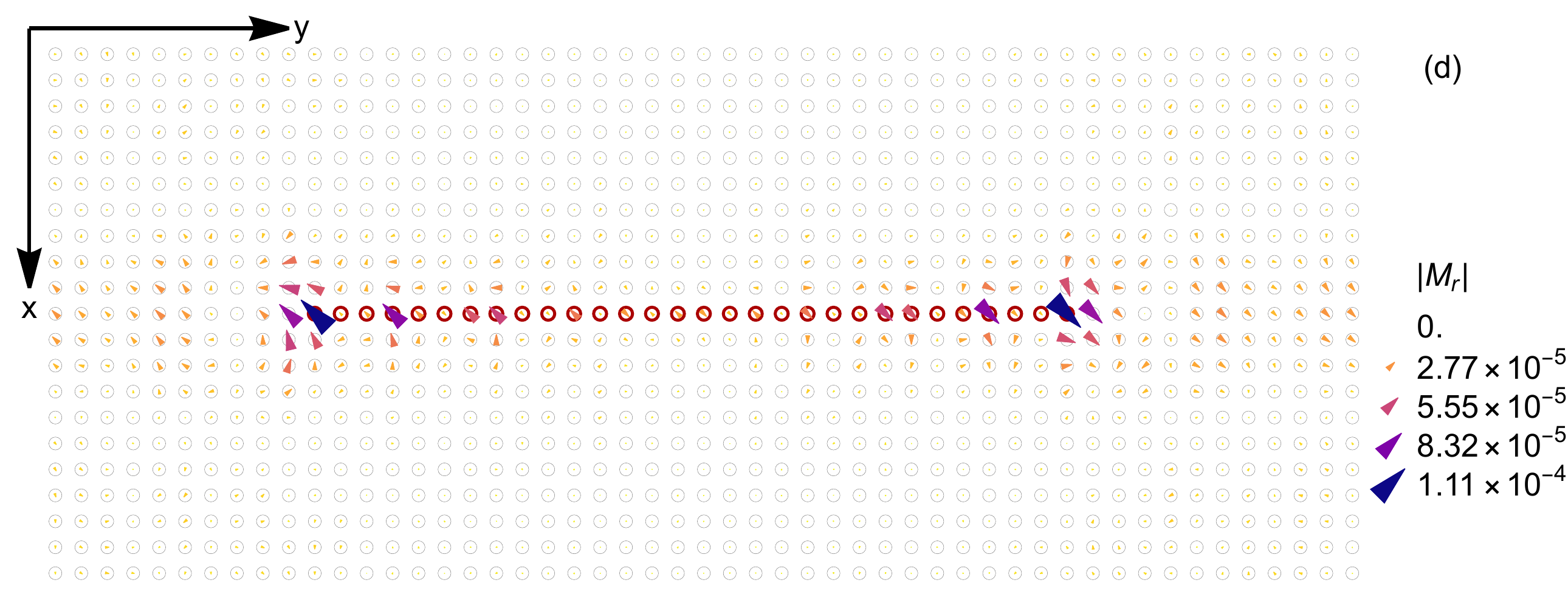}
  \caption{The Majorana polarization for a point in the phase space for which the substrate is topological (a) $U=2t$, $\kappa=0.2t$ and b) $U=3t$, $\kappa=0.2t$), non-topological close to the phase transition (c) $U=2t$, $\kappa=0.1t$) and far from it (d) $U=1t$, $\kappa=0.02t$). We take $\Delta= 0.16t$ and $\mu=3.5t$. Note that the Majorana polarization vectors of the lowest energy states are leaking in the bulk for a) and b), are concentrated mainly on the chain for c), and there is negligible Majorana character for d). The system size is $21\times 70$ (a,b) or $61\times 70$ (c,d), with a gap of $20$ sites between the boundary of the system and the end of the chain, periodic boundary conditions are imposed.
}\label{mpm4}
\end{figure}

In Fig.~\ref{mpm4}  we show plots of the local Majorana polarisation. When the substrate is topological the impurity states leak into the substrate and hybridise with the bulk states, see Fig.~\ref{mpm4}(a) and (b). The spatial profiles are similar to those obtained for magnetic impurities. When we place our parameters in the finely tuned region close to the topological bulk transition we observe better localized and polarized Majorana states, seeFig.~\ref{mpm4}(c). When the substrate is trivial, far from the topological transition, the states arising on the wire have negligible Majorana polarization, see Fig.~\ref{mpm4}(d).

\begin{figure}
 \includegraphics[height=0.4\columnwidth]{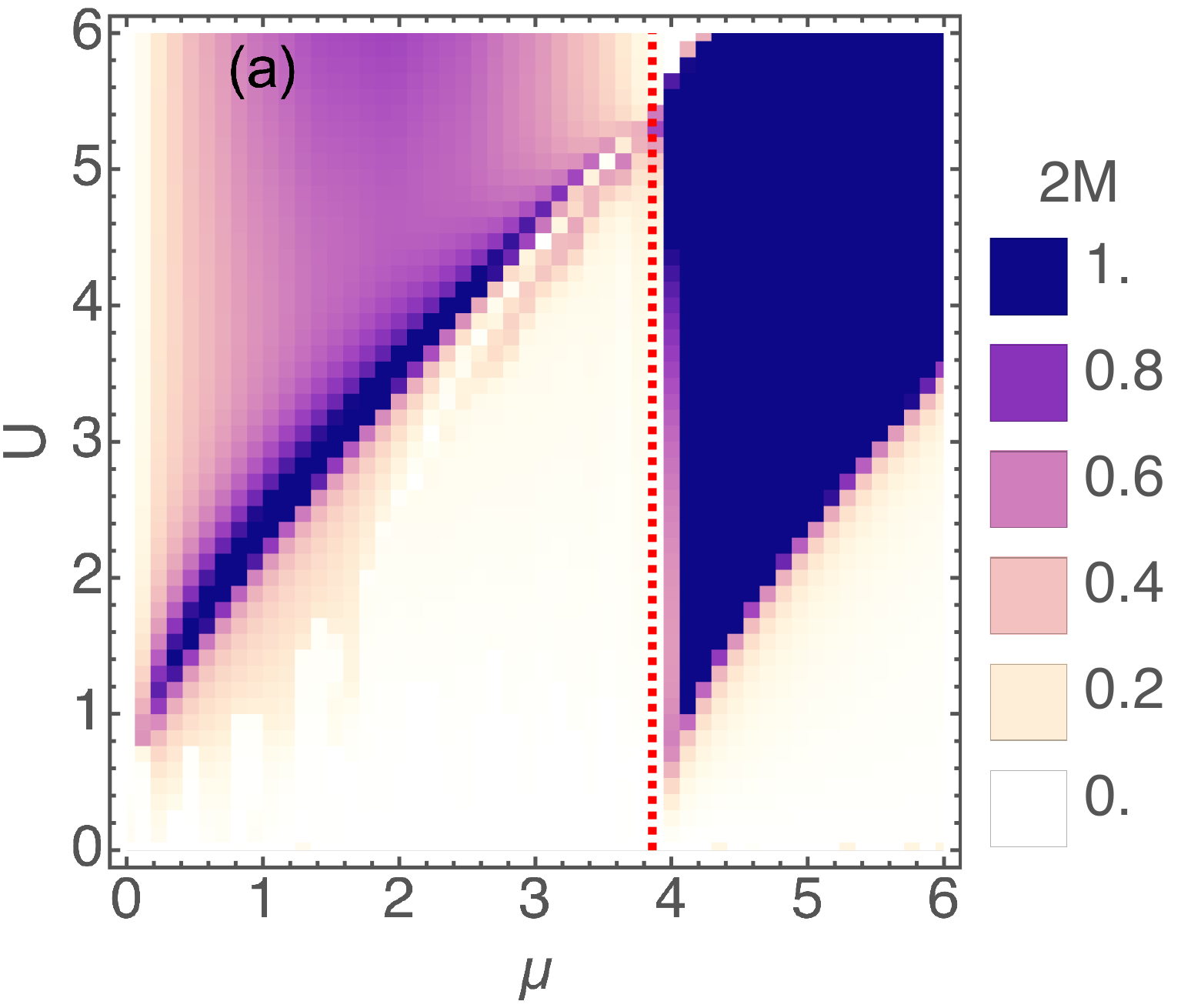}
 \includegraphics[height=0.4\columnwidth]{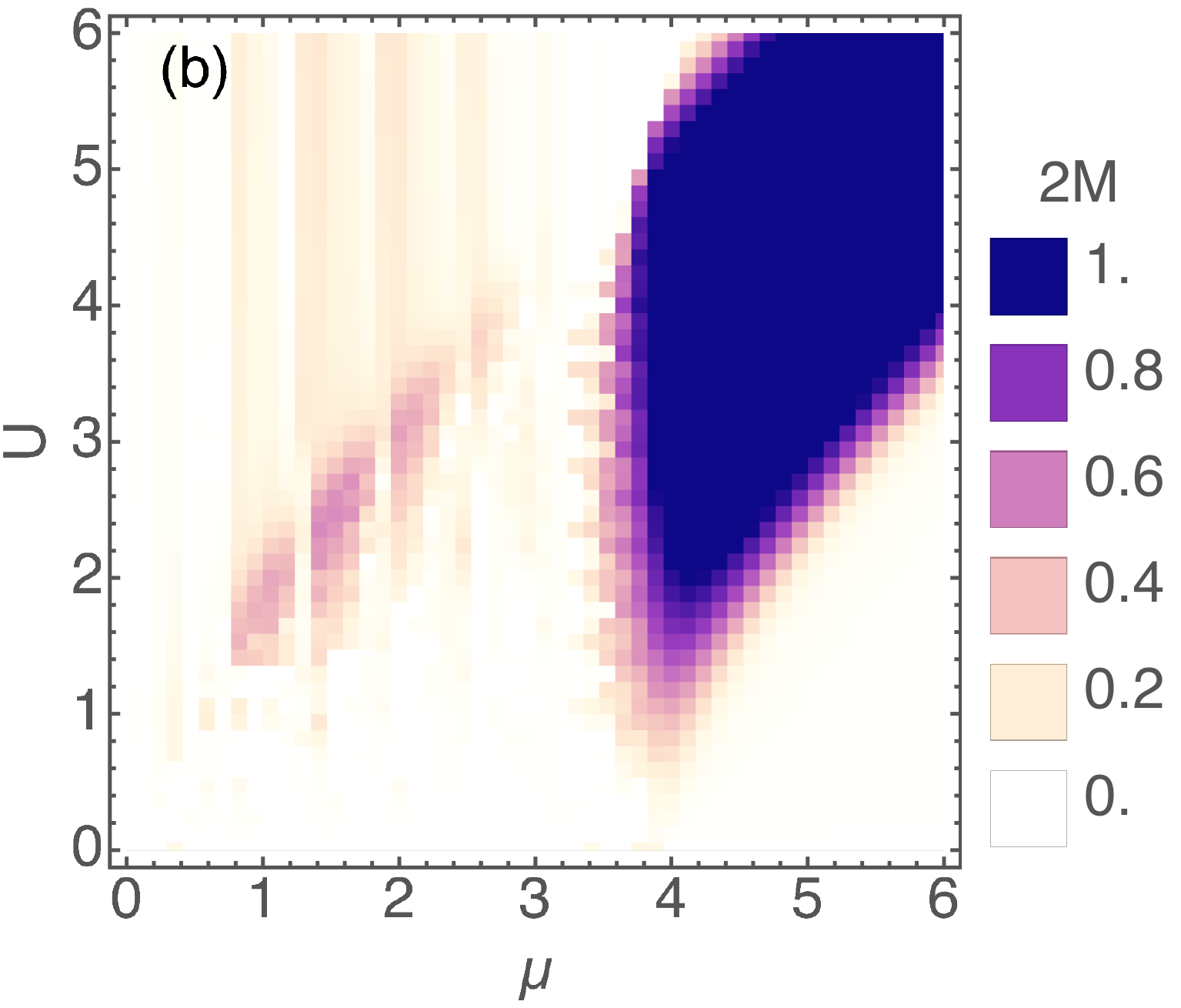}
  \caption{The Majorana polarization as a function of $\mu$ and $U$ in the bulk topological phase $\kappa=0.3t$ (a) and bulk non-topological phase close to the phase transition $\kappa=0.1t$ (b).  The system size is $21\times 80$ with a gap of $20$ sites between the impurities and the system end and periodic boundary conditions. The red dashed line is the topological phase boundary $2\kappa^2(4-\mu)=\Delta^2$ of the substrate. For (b) there is no non-trivial phase in the substrate as $\mu>4$.}
  \label{muU}
\end{figure}

We also plot the Majorana polarization phase diagram as a function of the chemical potential $\mu$ and the scalar impurity potential $U$ (see Fig.~\ref{muU}) for an s-wave component $\Delta=0.16t$ and two values of the p-wave parameter putting the substrate for $\mu<4-\Delta^2/2\kappa^2$ in the topological phase ($\kappa=0.3t$), as well as in the non-topological phase close to the transition ($\kappa=0.1t$). We note that, same as for the magnetic impurity case, for a large value of $\kappa>\Delta/2\sqrt{4-\mu}$ the the chain becomes topological only when the when the substrate is not, however when $\kappa$ is slightly  smaller than $\Delta/2\sqrt{4-\mu}$ we recover a slightly more extended topological phase diagram.

\begin{figure}
 \includegraphics[width=0.9\columnwidth]{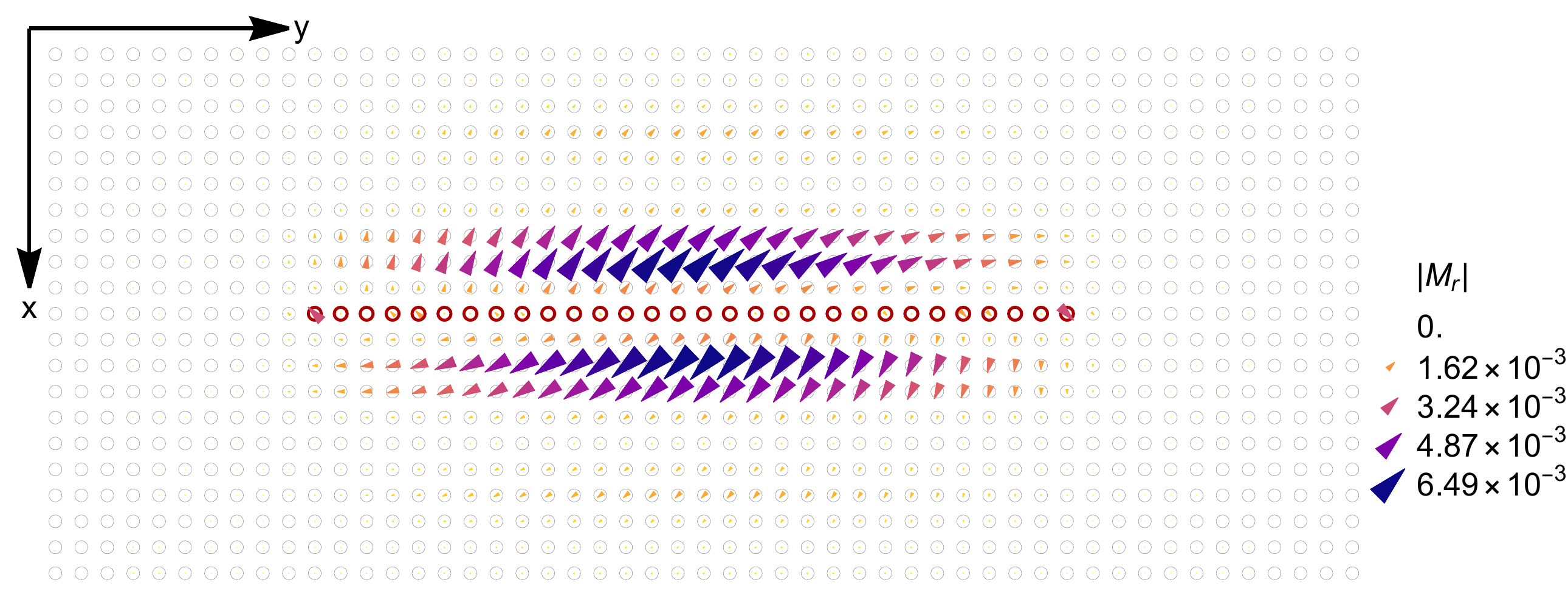}
  \caption{The Majorana polarization for a point in the phase space for a pure p-wave substrate. We take $U=3t$, $\Delta= 0t$, $\mu=3.5t$, and $\kappa=0.2t$.  The system size is $21\times 70$ with a gap of $20$ sites between the boundary of the system and the end of the chain, periodic boundary conditions are imposed.}
  \label{mpm5}
\end{figure}

Same as for the magnetic impurity we also consider the case of a pure p-wave, and in Fig.~\ref{mpm5} we plot the Majorana polarization as a function of position for a chain of scalar impurities. We note that the lowest-energy states in the wire also leak strongly in the substrate. Thus we conclude that clean and isolated Majorana states cannot be obtained on a topological substrate by depositing scalar impurities on a p-wave substrate, and we confirm our previous observation that it is quasi-impossible to recover clean Majorana states in a chain of impurities deposited on a topological substrate.

\section{Conclusion}
We have revisited the problem of chains of scalar impurities deposited on a superconducting substrate with mixed p-wave and s-wave components. We have also explored a similar configuration involving magnetic-impurity chains. For the magnetic impurity chains we have found that when the substrate itself is topological the Majorana states are not fully localized in the chains, they leak in the bulk and hybridize with each other, as well as with the bulk states, thus not making good quantum qubit candidates. This happens for both purely p-wave substrates, as well as for mixed s-wave/p-wave superconductors when the p-wave component is dominant. On the other hand when the substrate is trivial, i.e the s-wave component is dominant, the magnetic impurity chains can sustain clean Majorana states, well localized on the wire. The formation of Majorana states is quite a generic and stable phenomenon, and it is preserved for all the values of the p-wave and s-wave components, as long as the s-wave component is dominant and the p-wave component is non-zero. We note that here that the small p-wave component plays the same role as the Rashba spin-orbit coupling in the formation of Majoranas in chains of magnetic impurities on s-wave superconductors with spin-orbit.  For scalar chains it is much harder to obtain localized Majorana states: when the substrate is topological the Majorana states leak in the bulk and hybridize, while in the regime of a non-topological s-wave dominated substrate the chain can become topological only in a finely-tuned parameter regime close to the topological phase transition of the substrate. Our results indicate that a pure p-wave substrate would not be a good candidate to support topological Shiba chains for either magnetic or scalar impurities. However any small p-wave component in a mixed s-wave/p-wave superconductor would make possible the formation of Majoranas for a chain of magnetic impurities, while for a scalar impurity chain this would be much harder and require a significant amount of fine tuning. Our analysis was based in particular on the calculation of the local Majorana polarization and the density of the lowest-energy states forming in the system. 

\acknowledgments

This work was supported by the National Science Centre (NCN, Poland) under the grant 2019/35/B/ST3/03625 (NS). We would like to thank Vardan Kaladzhyan for interesting discussions.

\appendix

\section{Energy Scaling and Boundary Conditions}\label{app_energy}

\begin{figure}
	\includegraphics[height=0.3\columnwidth]{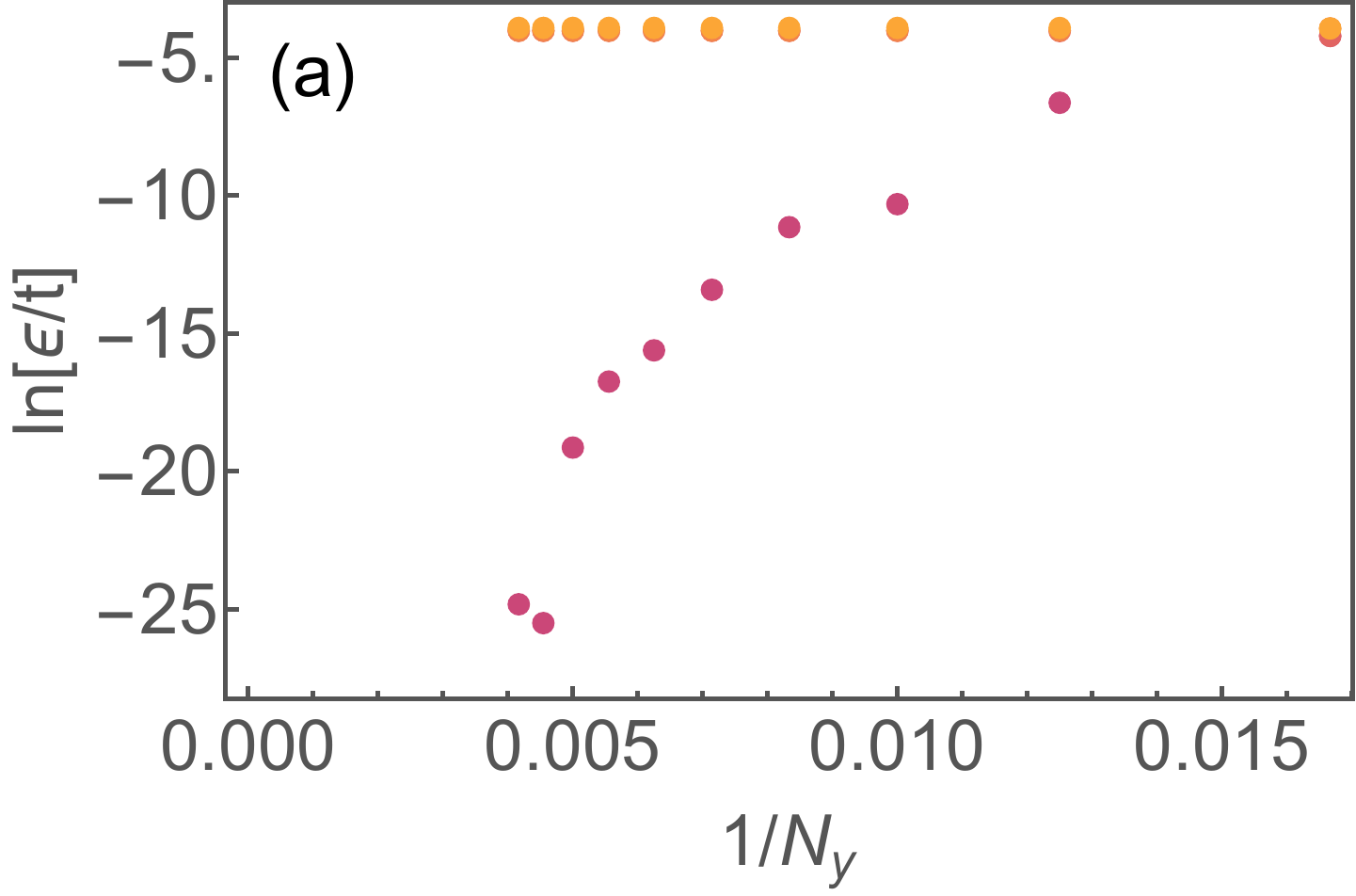}
	\includegraphics[height=0.3\columnwidth]{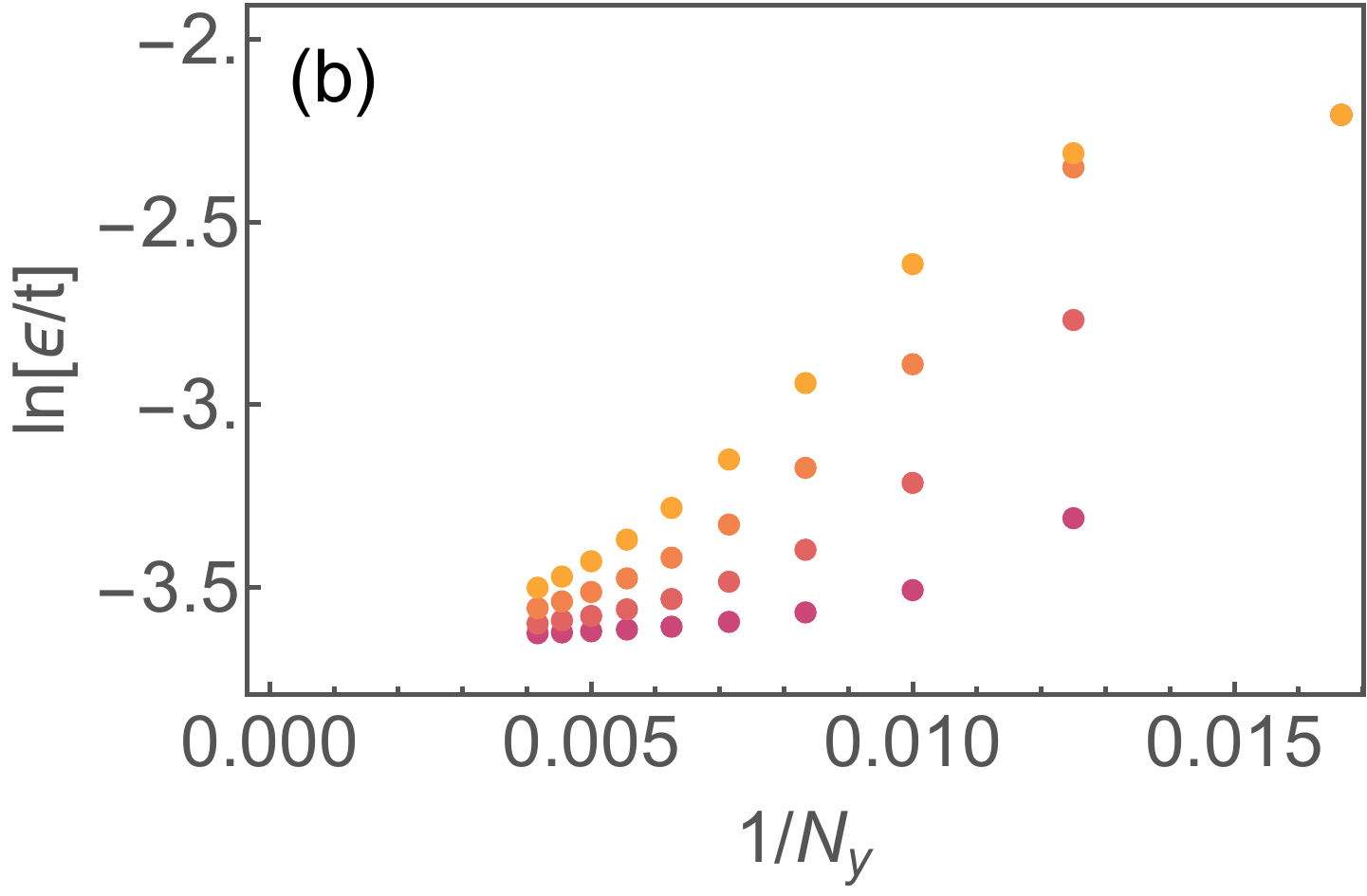}
  \caption{Lowest energies of the system with magnetic impurities as a function of inverse system length $N_y^{-1}$ for (a) open boundary conditions and (b) periodic boundary conditions along $x$. This is for $N_x=21$, $J = 3t$, $U = 0$, $\Delta = 0.16t$, $\mu = 3.5t$, and $\kappa= 0.2$, when the substrate is topologically non-trivial and $\nu_{\rm Q1D}$ is also non-trivial.}
  \label{app1}
\end{figure}

In this appendix we demonstrate that there is a fundamental difference between open boundary conditions and periodic boundary conditions for the zero modes. In Fig.~\ref{app1} we show the energy scaling of the lowest eigenvalues for a system with periodic and open boundary conditions in a regime which has a non-trivial quasi 1D invariant. As expected this system has an exponentially small eigenvalue. However for periodic boundary conditions in the lateral direction there is no zero mode.

\newpage

\section{Additional phase diagrams}\label{app_pd}

\begin{figure}
\vspace{0.2in}
	\includegraphics[height=0.4\columnwidth]{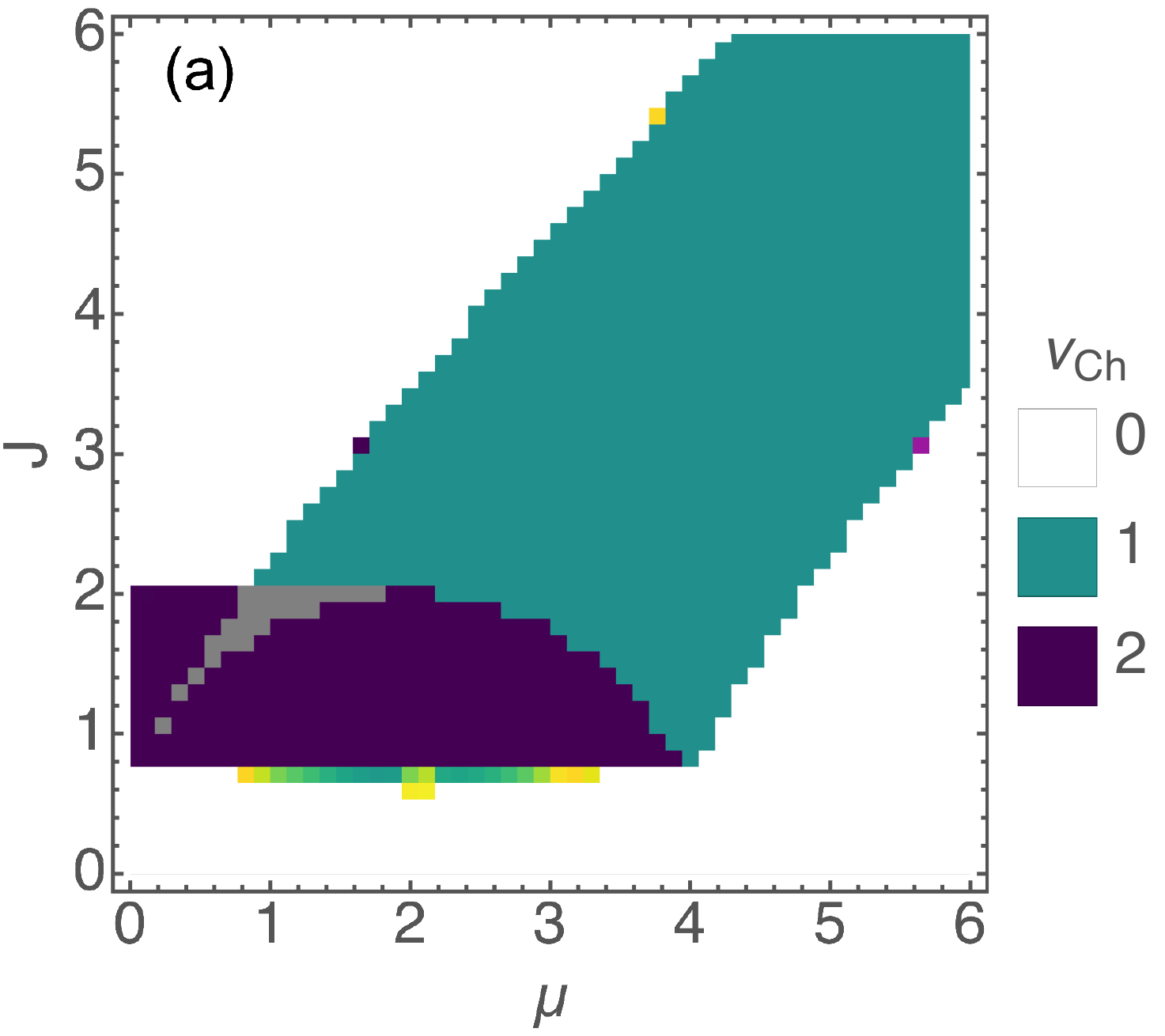}
	\includegraphics[height=0.4\columnwidth]{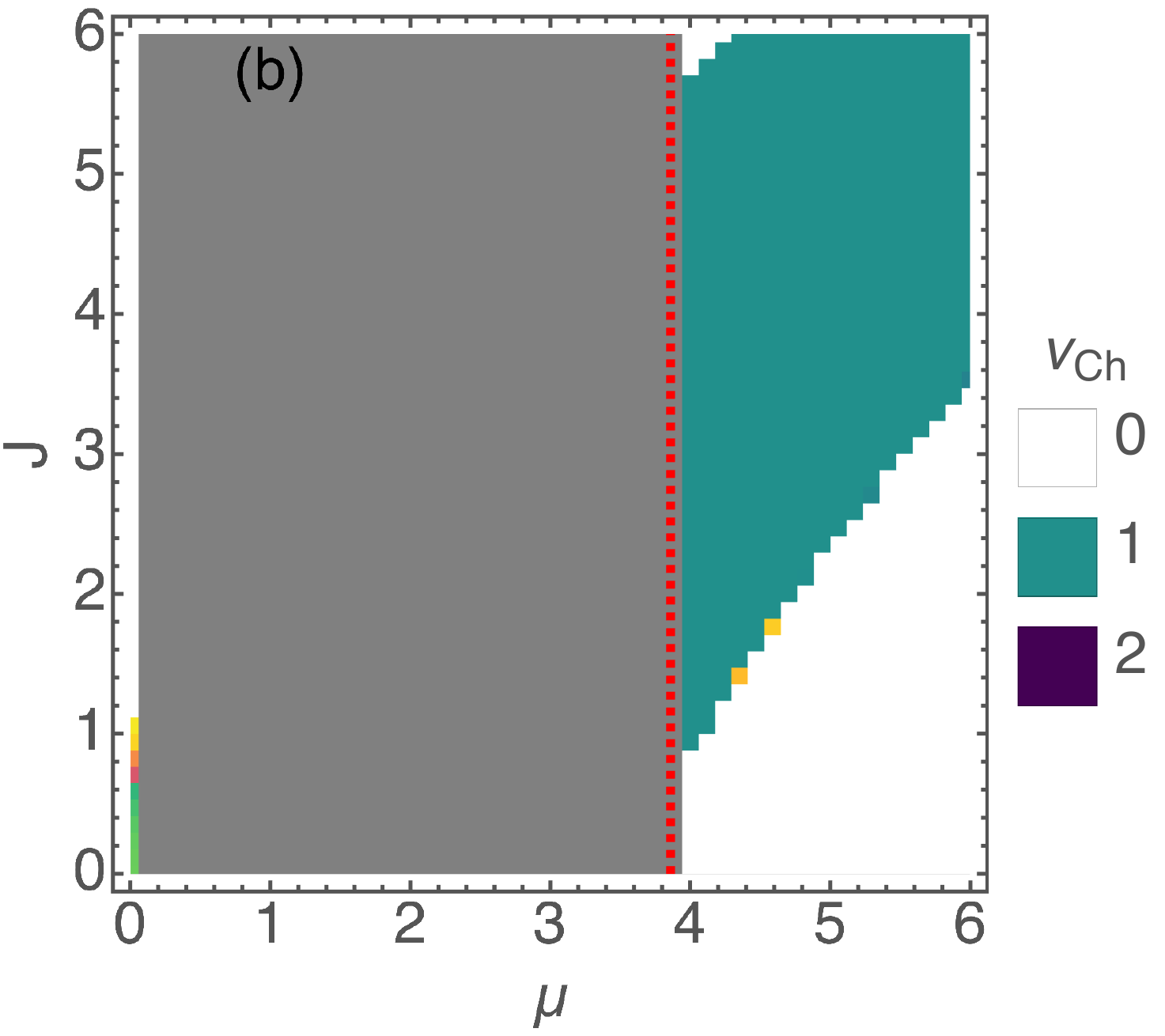}
  \caption{The chiral invariant as a function of $\mu$ and $J$, to compare with Fig.~\ref{muJ}. Here $\Delta= 0.16t$ with  $\kappa=0.05t$ (a) and  $\kappa=0.3t$ (b). The red dashed line is the topological phase boundary $2\kappa^2(4-\mu)=\Delta^2$ of the substrate. For (a) there is no non-trivial phase in the substrate as $\mu>4$. The grey region indicates the points where the integral for the chiral invariant did not converge in the available time. }
  \label{app2}
\end{figure}

Here we present additional phase diagrams calculated using the chiral invariant, see Fig.~\ref{app2}, to be compared to Fig.~\ref{muJ}. As before, when the substrate is topologically non-trivial the numerical integral over $k$ does not converge with the used number of steps in the Riemann sum.\vfill

%\bibliography{library}

%merlin.mbs apsrev4-1.bst 2010-07-25 4.21a (PWD, AO, DPC) hacked
%Control: key (0)
%Control: author (0) dotless jnrlst
%Control: editor formatted (1) identically to author
%Control: production of article title (0) allowed
%Control: page (1) range
%Control: year (0) verbatim
%Control: production of eprint (0) enabled
%

\end{document}